\documentclass[12pt]{amsart} 
\usepackage[margin=1.3in]{geometry}
\usepackage{graphicx}
\usepackage{multirow}
\usepackage[table,xcdraw]{xcolor}
\usepackage{graphicx,natbib}
\usepackage{amsmath}
\usepackage{placeins}
\usepackage{amssymb}
\usepackage{epstopdf}
\usepackage{pdfpages}
\usepackage{amsthm}
\usepackage{xcolor}
\usepackage{setspace}
\usepackage{amsaddr}

\newtheorem{theorem}{Theorem}[section]

\newtheorem{proposition}[theorem]{Proposition}

\newtheorem{example}[]{Example}[section]

\newtheorem{remark}[theorem]{Remark}

\newcommand*\E{\mathop{}\!\mathbb{E}}
\newcommand*\Ei{\mathop{}\!\mathtt{E}}
\newcommand*\V{\mathop{}\!\mathbb{V}}
\newcommand*\dd{\mathop{}\!\mathrm{d}}

\usepackage{multicol}

\title{Threshold selection and trimming in extremes}
\date{\today}

\author[M. \smash{Bladt}]{Martin Bladt (\MakeLowercase{corresponding author})}
\address[M. Bladt]{Department of Actuarial Science, Faculty of Business and Economics, University of Lausanne, CH-1015 Lausanne, Switzerland}

\email{{martin.bladt@unil.ch}}

\author[H. \smash{Albrecher}]{Hansj\"org Albrecher}
\address[H. Albrecher]{ Department of Actuarial Science, Faculty of Business and Economics and Swiss Finance Institute, University of Lausanne, CH-1015 Lausanne, Switzerland}

\email{{hansjoerg.albrecher@unil.ch}}

\author[J. \smash{Beirlant}]{Jan Beirlant}
\address[J. Beirlant]{Department of Mathematics,
	KU Leuven, Celestijnenlaan 200B, B-3001 Leuven, Belgium, and Department of Mathematical Statistics and Actuarial Science, University of the Free State, South Africa}

\email{{jan.beirlant@kuleuven.be}}

\begin{document}
\maketitle
\begin{abstract}
We consider removing lower order statistics from the classical Hill estimator in extreme value statistics, and compensating for it by rescaling the remaining terms. \color{black}{Trajectories of these trimmed statistics as a function of the extent of trimming turn out to be quite flat near the optimal threshold value.} 
For the regularly varying case, the classical threshold selection problem in tail estimation is then revisited, both visually via trimmed Hill plots and, for the Hall class, also mathematically via minimizing the expected empirical variance.  This leads to a simple threshold selection procedure for the classical Hill estimator {\color{black} which circumvents the estimation of some of the tail characteristics, a problem which is usually the bottleneck in threshold selection.} {\color{black} As a by-product, we derive an} alternative estimator of the tail index, which assigns more weight to large observations, and works particularly well for relatively lighter tails. A simple ratio statistic routine is suggested to evaluate the goodness of the implied selection of the threshold. We illustrate the {\color{black} favourable} performance and {\color{black}the} potential of the proposed method with simulation studies and real insurance data.
\end{abstract} 

\section{Introduction}
%
The use of Pareto-type tails has been shown to be important in different areas of risk management, such as for instance in computer science, insurance and finance. In social sciences and linguistics the model is referred to as Zipf's law. This model corresponds to the max-domain of attraction of a generalized extreme value distribution with a positive extreme value index (EVI) $\xi$:
\begin{equation}
1-F(x) = x^{-1/\xi}\ell (x), \quad \xi>0,
\label{Patype}
\end{equation}
where $\ell$ denotes a slowly varying function at infinity:
\begin{equation}
\lim_{x \to \infty}\frac{\ell (ux)}{\ell (x)}=1, \mbox{ for every } u>0. 
\label{ell}
\end{equation}
Since the appearance of the paper of \cite{Hill} in which the EVI estimator 
\begin{equation}
H_{k,n} = {1 \over k}\sum_{i=1}^k \log X_{n-i+1,n}-\log X_{n-k,n} 
\label{Hill}
\end{equation}
was proposed with 
$$
X_{n,n}  \geq X_{n-1,n} \geq \cdots \geq X_{n-i+1,n}\geq \cdots \geq X_{1,n}
$$
denoting the ordered statistics of a random sample from $F$, the literature on estimation of $\xi>0$ and other tail quantities  such as extreme quantiles and tail probabilities has increased exponentially. We refer to \cite{embrechts2013modelling}, \cite{beirlant2006}, \cite{de2007extreme} and \cite{gomes2015extreme} for detailed discussions and reviews of these estimation problems. Next to the proposal of numerous estimators, focus has gradually shifted to selection methods of $k$ and to the construction of bias-reduced estimators which exhibit plots of estimates which, as a function of $k$, are as stable as possible. Indeed, plots of estimators of $\xi$ as a function of $k$ that are consistent under the large semi-parametric model \eqref{Patype} are hard to interpret. In case of the Hill estimator some authors refer to Hill horror plots. While it has been frequently suggested to choose a 'stable' area (see for instance \cite{drees2000make} and \cite{de2004diagnostic}), such a stable part is often absent or hard to find. Sometimes more than one stable section is present, like in some insurance applications as we will discuss later.\\

\noindent
The typical available guidelines for the choice of $k$ to be used in the implementation of the EVI estimators depend strongly on the properties of the tail itself, and $k$ needs to be estimated adaptively from the data. This problem can be compared with choosing a bandwidth parameter in density estimation. It is typically suggested that the optimal value of $k$ should be the one that minimizes the mean-squared error (MSE). However, this optimum depends on the sample size, the unknown value of $\xi$ as well as on the nature of { the slowly varying function} $\ell$, as was first described in \cite{hall1985adaptive}.  
Bootstrap methods were proposed in \cite{hall1990using}, \cite{draisma1999bootstrap}, \cite{danielsson2001using}, and \cite{gomes2001bootstrap}. \cite{beirlant1996tail,beirlant2002exponential} derived regression diagnostic methods on a Pareto quantile plot.
Other selection procedures can be found in \cite{drees1998selecting} and \cite{guillou2001diagnostic}. Possible heuristic choices are provided in \cite{gomes2007sturdy}, \cite{gomes2008tail} and \cite{beirlant2011generalized}.  {Recent proposals rooted in goodness-of-fit approaches are found in \cite{Bader}, \cite{Drees_Janssen} and \cite{Schneider}. } 
Almost all authors consider the adaptive choice of $k$ for the Hill estimator.\\

\noindent
In this paper we consider trimming of the Hill estimator, omitting some of the lower order statistics in 
$ 
X_{n-k+1,n},\ldots, X_{n,n},
$
which leads to \color{black}{statistics} of the type
\begin{equation}\label{def1}
T_{b,k}=\sum_{i=1}^b c_i (b,k) \log\left( {X_{n-i+1,n}\over X_{n-k,n}}\right),
\end{equation}
for some $1 \leq b \leq k$ and suitable constants $c_i (b,k)$. {\color{black} This kind of kernel-type statistics have been previously proposed  (cf.  \cite{CsDM})} \color{black}{as estimators of $\xi$}.  \color{black}{However,} \color{black}{the implementation of the optimal kernel is not an easy task nor our focus in this paper. Instead, we propose a special form of the kernel \color{black}{that leads to an identity which} aids in the threshold estimation problem.} In Section 2 we \color{black}{derive the coefficients $c_i (b,k)$ \color{black}{which} make $T_{b,k}$ unbiased when $\ell$ is constant and when we  force  the coefficients $c_i (b,k)= c(b,k)$  not to depend on $i$}. We present a novel lower-trimmed Hill plot which provides significant graphical support for the estimation problem of $\xi$, as we illustrate with both simulations and real world data. \color{black}{We also provide mathematical evidence that, as a function of $b$, the variability of the $T_{b,k}$ statistics is lower \color{black}{than the one} in the Hill plot}. 
 In Section 3, we examine the asymptotic characteristics of \color{black}{$T_{b,k}$} in \eqref{def1} under the general model \eqref{Patype}. The asymptotic expected empirical variance of \color{black}{$T_{b,k}$} is shown to be less sensitive on the tail parameter $\xi$ than the asymptotic mean-squared error (AMSE) of the usual Hill estimator \eqref{Hill}. We identify a link between \color{black}{the corresponding two optimal  $k$-choices}   which allows to bypass the specification of $\xi$ and other characteristics of the tail behavior for the identification of the optimal threshold in the classical Hill estimate, and the resulting procedure turns out to be simple to implement in practice.  
 Subsequently, we study the estimator $\overline{T}_k$ obtained by averaging the trimmed Hill estimators over $b=1,\ldots,k$. This latter estimator naturally assigns more weight to the larger observations, the weights being only moderately changed when increasing $k$. Furthermore, the specification of these weights is independent of the distribution $F$. 
 Note that, in contrast,  earlier criteria for  reweighting terms in the Hill estimator (such as e.g.\ \cite{CsDM} in terms of kernel estimates, see also \cite[Sec.3]{beirlant2002exponential}) had to heavily rely on the tail parameter $\xi$. 
 In Section 4 we then present a simple ratio statistic as a tool to evaluate the goodness of selection of $k$. Section 5 confirms the good performance of the proposed methods using simulations, where $\overline{T}_k$ turns out to outperform the classical Hill estimator in almost all cases. Note that our approach eventually suggests a fully automated procedure for the threshold selection, also in the absence of knowledge about, or assumptions on, the tail characteristics. Section 6 favorably illustrates this on a set of real-life motor third party liability insurance data. We would like to emphasize that the approach proposed in this paper suggests a general procedure that can in principle also be applied to other estimators in extreme value analysis. 

\section{A lower-trimmed Hill statistic}
\subsection{Derivation}
Assume first, for simplicity, that we have independent and identically distributed (i.i.d.) exact Pareto random variables, $X_1,X_2,\dots,X_n$, with tail given by
\begin{align}\label{exactparetomodel}
\overline F(x)=(x/\sigma)^{-1/\xi}, \quad x\ge \sigma, \quad \xi,\sigma>0,
\end{align}
and we are interested in robust estimation of the tail index $\xi$. 

A main tool used throughout the paper is the well-known \textit{R\'{e}nyi representation}, which states (in the second distribution equality below), that for the order statistics of a random sample $X_1,\dots, X_n$ from the distribution \eqref{exactparetomodel}, one has, for $k\le n$, with $Y_{i,k}= X_{n-i+1,n}/X_{n-k,n}$ $(i=1,\ldots,k)$, 
\begin{align}
&\left(\log\left(Y_{1,k} \right),\dots,\log\left( Y_{k,k}\right)\right)\stackrel{d}{=}(E_{k,k},\dots,E_{1,k})\stackrel{d}{=} \left(\sum_{j=1}^k\frac{E_j^\ast}{k-j+1},\dots,\frac{E_1^\ast}{k}\right).\label{Renyi_rep}
\end{align}
Here, $E_{k,k}\ge \cdots \ge E_{1,k}$ are the order statistics of an independent i.i.d.\ exponential sample $E_1,\dots, E_k$ with mean $\xi$, and $E_1^\ast,\dots,E_k^\ast$ is another independent i.i.d.\  exponential sample with mean $\xi$. \\

\cite{hilltrim} recently proposed linear estimators of the form 
\begin{align*}
\hat\xi_{k_0,k}=\sum_{i=k_0+1}^k c_{k_0,k}(i)\log\left( Y_{i,k}\right),\quad 0\le k_0<k<n,
\end{align*}
in order to trim the upper order statistics in outlier-contaminated samples, where the constants $c_{k_0,k}(i)$ are chosen in a way to ensure that the resulting estimator for $\xi$ is unbiased. 
For fixed $k_0,k,$ the problem can then be recast into that of finding suitable weights $\delta_i$ such that one can write
\begin{align*}
\hat\xi_{k_0,k}=\sum_{i=k_0+1}^k c_{k_0,k}(i)E_{k-i+1,k}=\sum_{i=1}^{k-k_0}\delta_iE_{i,k}.
\end{align*}
Using the R\'{e}nyi representation \eqref{Renyi_rep} and solving some elementary linear equations, they derived $\delta_i=\frac 1 r, \; i<r,$ and $\delta_r=({k-r+1})/{r}$. This led them to the so-called \textit{trimmed Hill estimator}
\begin{align*}
\hat\xi_{k_0,k}&=\frac{k_0+1}{k-k_0}\log\left(Y_{k_0+1,k}\right)+\frac{1}{k-k_0}\sum_{i=k_0+2}^k\log\left(Y_{i,k}\right),
\end{align*}
which is shown to be quite useful in outlier detection under \eqref{Patype}.

In a similar way, but for a different purpose, in this paper we investigate trimming from the left. Concretely, we consider estimators of the form
\begin{align*}
T_{b,k}=\sum_{i=1}^b c_i(b,k)\log\left(Y_{i,k}\right),\quad 0 <b\le k,
\end{align*}
where $c_i(b,k)$ are constants to be determined. As above, we would like to find suitable weights $\gamma_i$ such that 
\begin{align}\label{generictbkform}
T_{b,k}=\sum_{i=1}^b c_i(b,k)E_{k-i+1,k}=\sum_{i=k-b+1}^{k}\gamma_iE_{i,k}
\end{align}
Setting $q=k-b+1$, the R\'{e}nyi representation \eqref{Renyi_rep} yields 
\begin{align*}
T_{b,k}=\sum_{i=q}^{k}\gamma_iE_{i,k}&=\sum_{i=q}^k \gamma_i \sum_{j=1}^i\frac{E_j^\ast}{k-j+1}\\
&=\sum_{j=1}^k E_j^\ast \sum_{i=j\vee q}^k\frac{\gamma_i}{k-j+1}=\sum_{j=1}^k \overline \gamma_j E_j^\ast 
\end{align*}
with $\overline \gamma_j:=\sum_{i=j\vee q}^k\frac{\gamma_i}{k-j+1}$. Here we use the notation $j\vee q=\max\{j,q\}$. Unfortunately, the set of equations 
\begin{align*}
\overline\gamma_j=\frac 1k, \quad j=1,\dots, k,
\end{align*}
has no solution (for $j\le q$ the left-hand-side cannot remain constant in $j$).
Instead, we choose to set 
\begin{align}\label{defeq1}
\gamma_q=\gamma_{q+1}=\cdots=\gamma_k=:\frac{1}{\overline\omega(q,k)}\end{align}
 and 
 \begin{align}\label{defeq2}
 \E\left[T_{b,k}\right]=\xi
 \end{align}
as the defining equations. The solution of \eqref{defeq1} and \eqref{defeq2} is given by
\begin{align}\label{omegabardef}
\overline\omega(q,k)=\sum_{j=1}^k\frac{k-j\vee q+1}{k-j+1}.
\end{align}
Plugging \eqref{omegabardef} into \eqref{generictbkform}, we then arrive at the following definition \color{black}{ of a \textit{lower-trimmed Hill statistic}} $T_{b,k}$:
\begin{align}
T_{b,k}&=\sum_{i=q}^k \frac{\log\left(Y_{k-i + 1,k}\right)}{\overline\omega(q,k)}=\frac{\sum_{i=1}^{b} \log\left(Y_{i,k}\right)}{\overline\omega(k-b+1,k)}\nonumber\\
&=\frac{{1 \over b}\sum_{i=1}^b\log(Y_{i,k})}{1+\sum_{j=b+1}^kj^{-1}}, \quad b=1,\dots,k, \:\: k< n,\label{lth}
\end{align}
where we use the convention $\sum_{j=k+1}^kj^{-1}:=0$.
\\
{ Note that $T_{b,k}$ can be considered as a trimmed pseudo-likelihood estimator of $\xi$ under the strict Pareto model \eqref{exactparetomodel}. Indeed, under  \eqref{exactparetomodel} the exceedances $Y_{i,k}$ ($i=1,\ldots,k$) are distributed as the order statistics from a sample of size $k$ from \eqref{exactparetomodel} with $\sigma=1$. Then the trimmed likelihood, considering the conditioning on the exceedances larger than $Y_{b+1,k}$, is given by
\begin{align*}
\Pi_{i=1}^b \left( {1 \over \xi}Y_{i,k}^{-1/\xi -1}\right) / Y_{b+1,k}^{-1/\xi}, 
\end{align*}
leading to the likelihood equation
\begin{align}
{1 \over \xi} \; {1 \over b}\sum_{i=1}^b \log Y_{i,k}
= 1+ \log Y_{b+1,k}^{1/\xi}.
\label{MLeq}
\end{align}
From \eqref{Renyi_rep} it follows that 
$
\mathbb{E}(\log Y_{b+1,k}^{1/\xi})= \sum_{j=b+1}^k j^{-1}
$, from which the estimator $T_{b,k}$ follows after formal substitution of $\log Y_{b+1,k}^{1/\xi}$ by its expected value in \eqref{MLeq}.
}

\subsection{A lower-trimmed Hill plot}

{ The so-called Hill plots, in which $T_{k,k}$ are plotted as a function of  $k$, probably are the most popular starting tool in extreme value analysis for Pareto-type tails. However, the difficulties involved when interpreting these plots and searching for 'horizontal' or 'stable' parts that indicate the start of a tail part that resembles a pure Pareto tail, if at all available, constitute a serious obstacle in practical applications.  In the literature, two lines of research have been developed in order to remedy these problems: adaptive selection of $k$ along some criteria such as minimization of the MSE, or bias-reduction searching for estimators that are more stable, hence reducing the drift due to bias when $k$ is taken too large. Concerning the line of research on bias reduction we can refer to recent papers using ridge regression methods  in \cite{ridge} or Mean-of-Order-$p$ estimators from \cite{mop}, and the many other proposals cited in those papers. Both approaches, adaptive selection and bias reduction, can provide extra insight in a given case and complement each other. For instance, with increasing $k$ bias-reduced estimators often start deviating from the original Hill plot at or around MSE-optimal $k$ levels.  Moreover, next to the selection of an appropriate threshold,  stable bias-reduced methods based on extended Pareto models can provide models that fit well on a larger set of top data, see for instance  \cite{Papa_Tawn} and the references therein. \\ 
}

{ While in the next sections we propose an adaptive selection method for $k$, we also suggest plotting  $T_{b,k}$ defined above for $b=1,\ldots,k$ and different $k$,} as it is unbiased for any $b,k$, $b\le k$ under \eqref{exactparetomodel} by construction. Analogous to the Hill plot,  we  exploit the second degree of freedom and plot, for selected values of $k$, $T_{b,k}$ as a function of $b$. That is, the plot is constructed by overlaying the trajectories
\begin{align*}
(b,\,T_{b,k}),\quad b=1,\dots, k,
\end{align*}
for a selection of $k$ values. The lower variance of these trajectories, comes from the fact that the normalizing order statistic is fixed, and hence a non-constant behaviour is easier to identify  visually than in the classical Hill plot. As a particular consequence, the selection of $k$ that makes the tail resemble a pure Pareto tail is easier to determine, by examining when the trajectories start to be constant,
{ and hence indicating zones with reduced bias.}
\\

The following Proposition provides mathematical evidence for the above observations.
\begin{proposition}
As a function of the number $b$ of order statistics being used, in the exact Pareto case \eqref{exactparetomodel} the estimator 
$T_{b,k}$ has lower variance than the classical Hill estimator $T_{b,b}$. More precisely,
\begin{align*}
\V\left[T_{b,b}\right]=\frac{\xi^2}{b}\quad\text{and}\quad 
\V\left[T_{b,k}\right]\le\frac{\xi^2}{\sum_{j=1}^{k-b+1}\left(\frac{b}{k-j+1}\right)^2+b}.
\end{align*}
\end{proposition}

As an illustration, we now compare the performance of
 these lower-trimmed Hill (LTH) plots for Pareto, near-Pareto and spliced Pareto distributions. The latter is defined through its cumulative distribution function (c.d.f.)
\begin{align}\label{slicedcdf}
F(x;\xi_0,r,c)=\frac{(1-x^{-1/\xi_0-r\cdot 1\{x\ge c\}})-1\{x\ge c\}(c^{-1/\xi_0}-c^{-1/\xi_0-r})}{1-c^{-1/\xi_0}+c^{-1/\xi_0-r}}, \;x\ge 1
\end{align}
for $c\ge 1$ and $r>-1/\xi_0$, which is the c.d.f.\ of a Pareto random variable with tail index $\xi_0$ up to some splicing point $c$, continuously pasted with the c.d.f. of a Pareto random variable with another tail index $\xi=(1/\xi_0+r)^{-1}$ thereafter. Splicing models (also sometimes referred to as composite models) are for instance popular in reinsurance modelling, cf.\  \cite[Ch.4]{abt}. \\

Concretely we simulated a sample of size $n=1000$ from a:

\begin{itemize}
\item pure Pareto $\xi=\sigma=1$ sample, defined in \eqref{exactparetomodel}.\\
\item spliced Pareto sample, defined in \eqref{slicedcdf}, with parameters $\xi=1$, $\xi_0=1/4$ and splicing point $c=1.3$.\\
\item Student-t distribution with $10$ degrees of freedom. The absolute value function was applied to the sample.\\
\item Burr sample with tail $\overline F(x)=\left(\frac{1}{1+x}\right)^5,\; x>0$ (which amounts to a generalized Pareto distribution).\\
\item Log-gamma with logshape parameter $3/2$ and lograte parameter $1$, i.e. with density $f(x)= {1  \over \Gamma(3/2) }\; (\log x)^{1/2} x^{-2},\;\; x>0$, with $\xi=1$.\\
\end{itemize}

\begin{figure}[hh]
\centering
\includegraphics[width=15cm,trim=.5cm .5cm .5cm .5cm,clip]{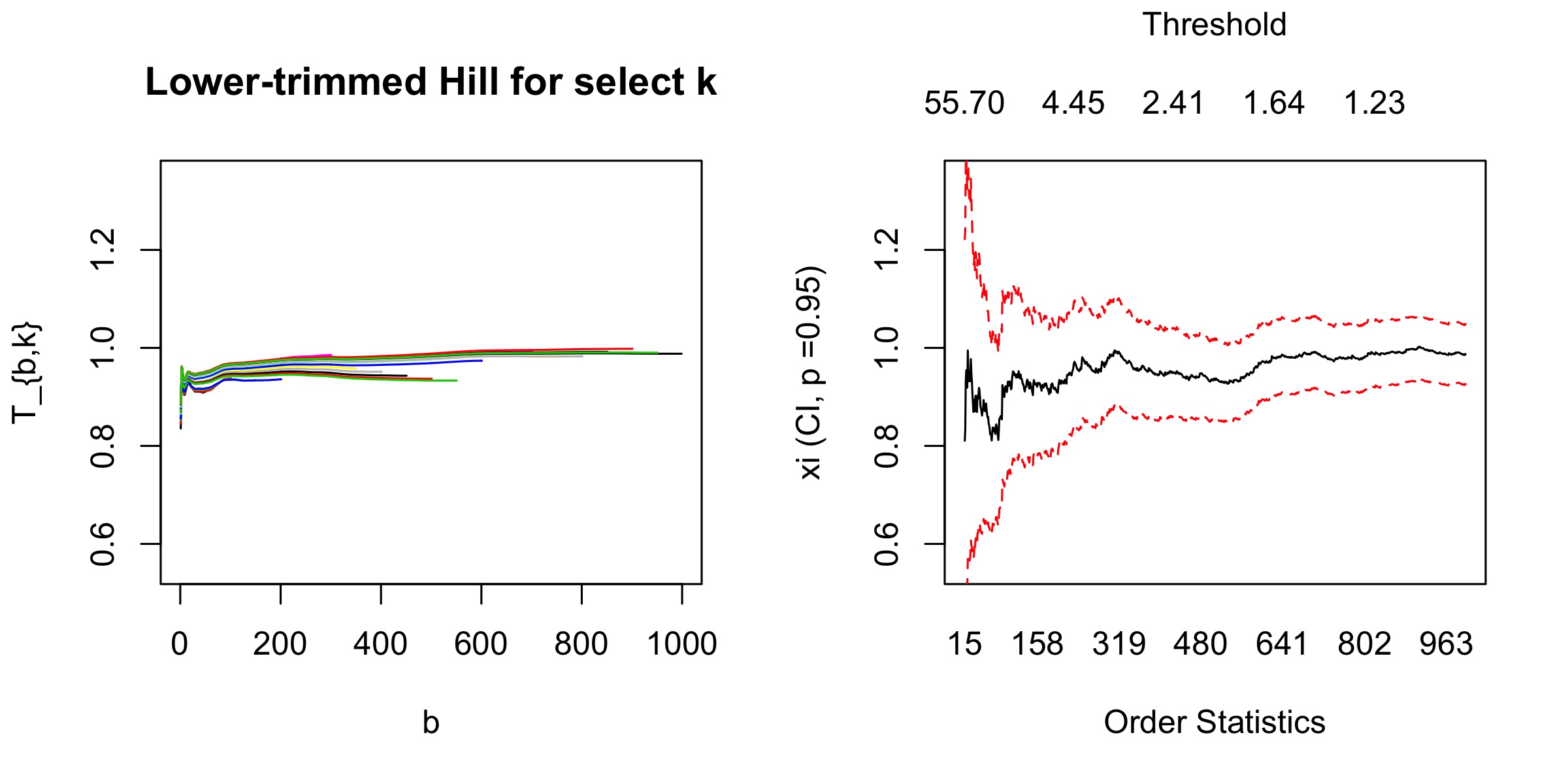}
\includegraphics[width=15cm,trim=.5cm .5cm .5cm 8cm,clip]{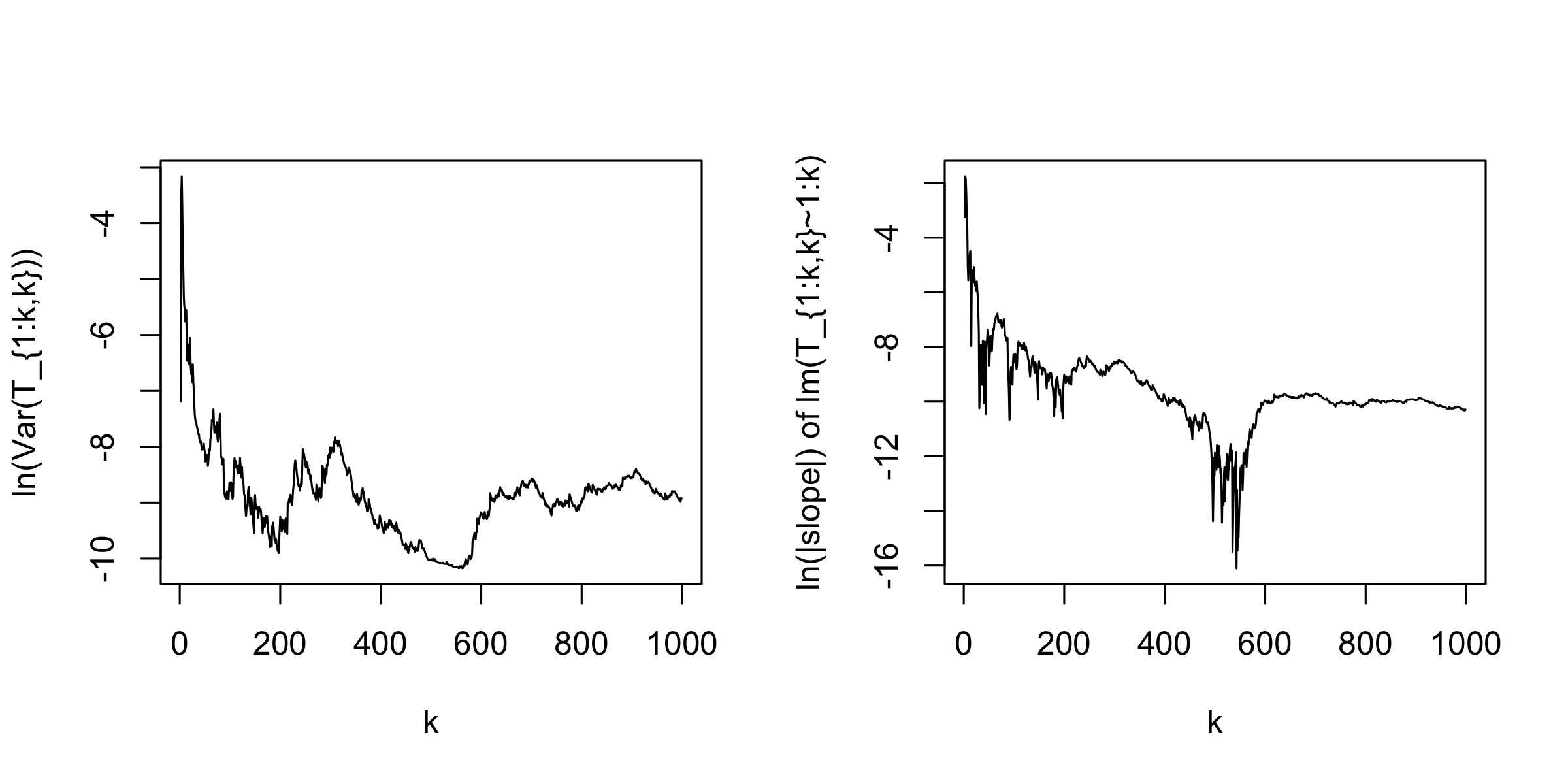}
\caption{Exact Pareto case ($\xi=1$). Top left: $T_{b,k}$ for varying lower trimming $b$, for $k=1,51,101,\ldots,1000$. Top right: Hill plot (black, solid). Bottom left: empirical variance of the LTH  as a function of $k$. Bottom right: slope of a fitted linear model to the LTH  as a function of $k$.} 
\label{dthe}
\end{figure}

\begin{figure}[hh]
\centering
\includegraphics[width=15cm,trim=.5cm .5cm 0cm .5cm,clip]{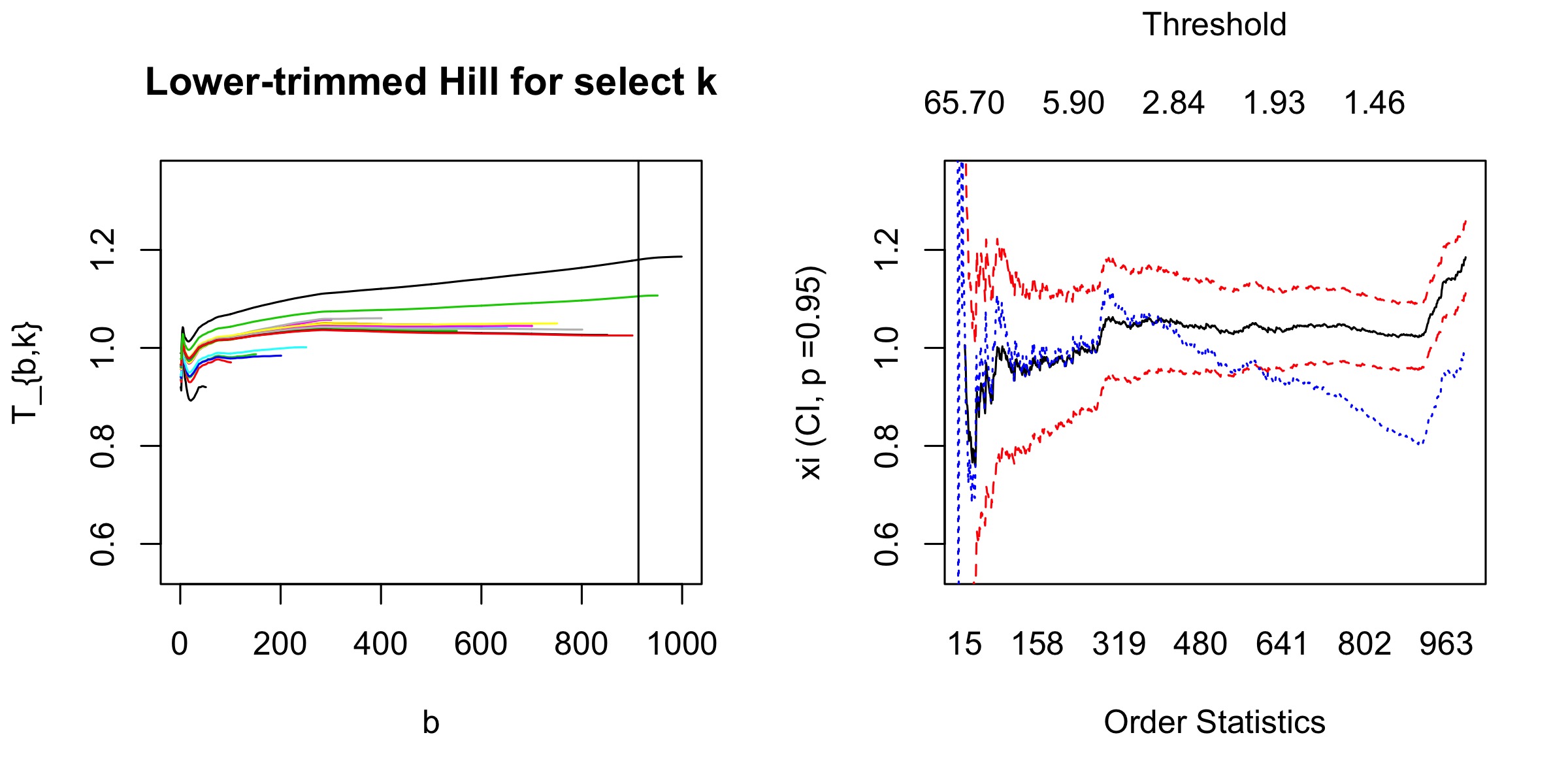}
\includegraphics[width=15cm,trim=.5cm .5cm .5cm 8cm,clip]{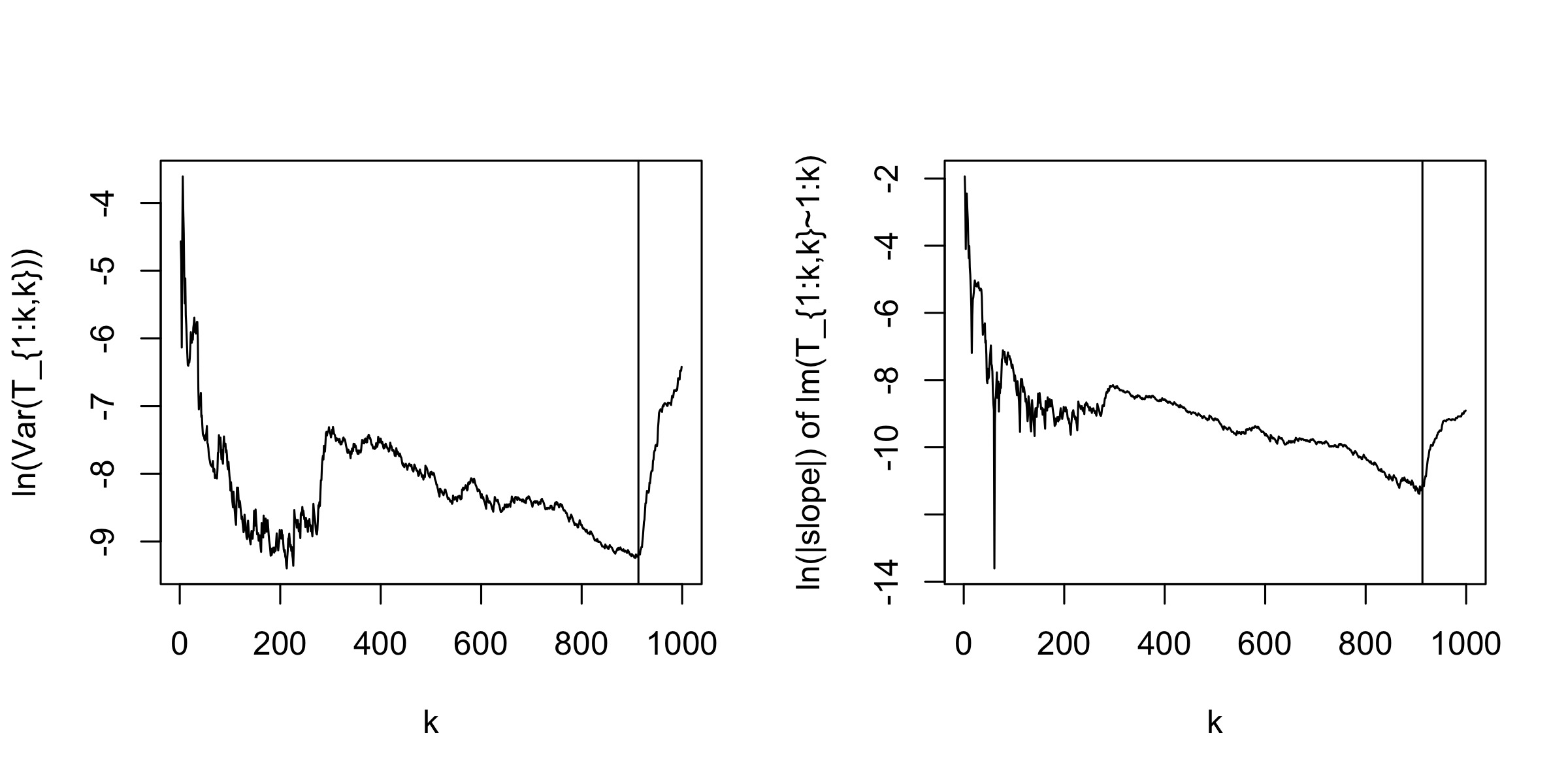}
\caption{Spliced Pareto case (body parameter: $1/4$ and tail parameter: $1$). Top left: $T_{b,k}$  for varying lower trimming $b$, for $k=1,51,101,\ldots,1000$. The vertical line is the splicing location. Top right: Hill plot { (black, solid), together with a mean-of-order $-1$ bias-reduced estimator (blue, dotted)}. Bottom left: empirical variance of the LTH  as a function of $k$. Bottom right: slope of a fitted linear model to the LTH  as a function of $k$.} 
\label{dths2}
\end{figure}

\begin{figure}[hh]
\centering
\includegraphics[width=15cm,trim=.5cm .5cm 0cm .5cm,clip]{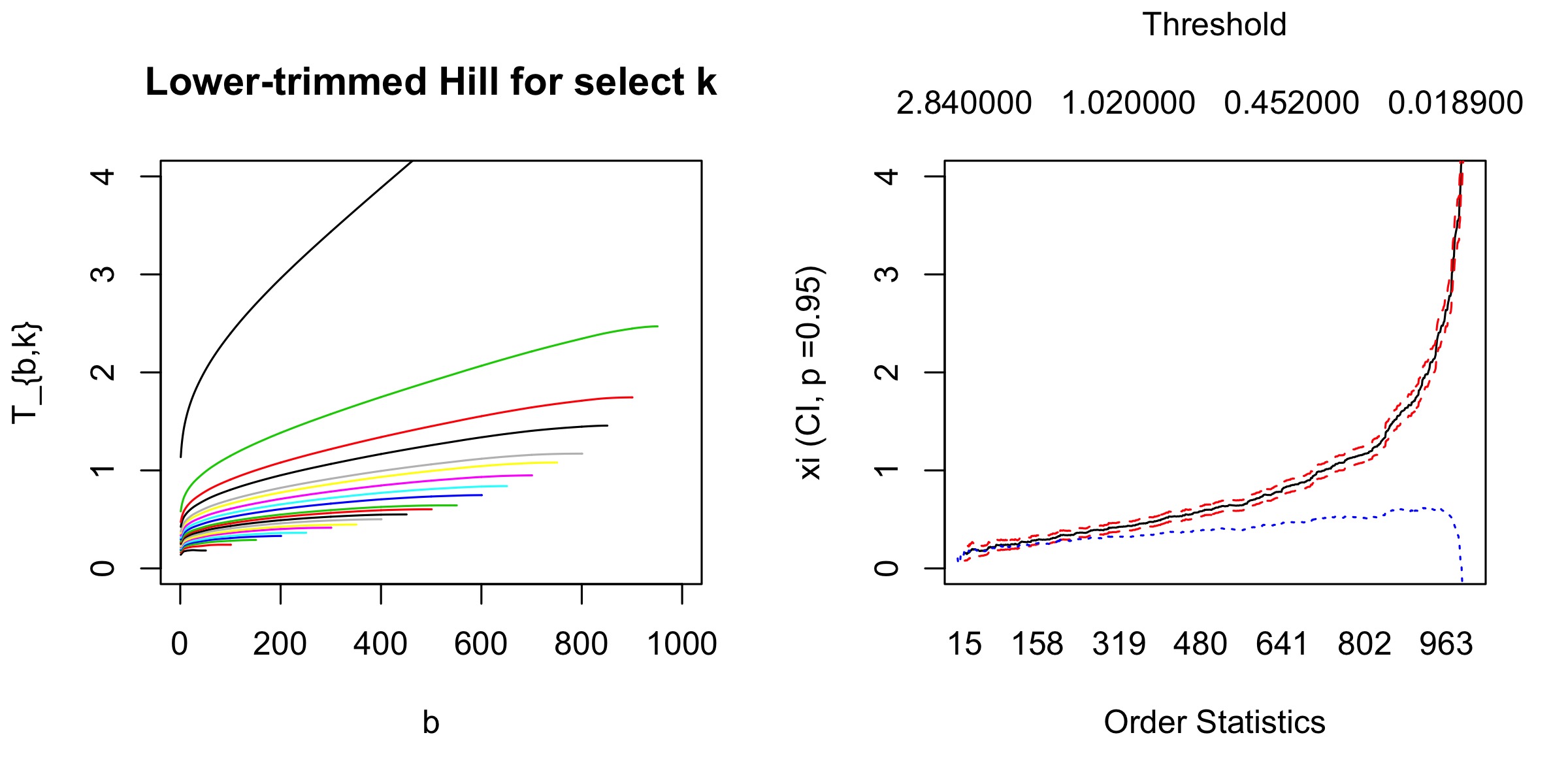}
\includegraphics[width=15cm,trim=.5cm .5cm .5cm 8cm,clip]{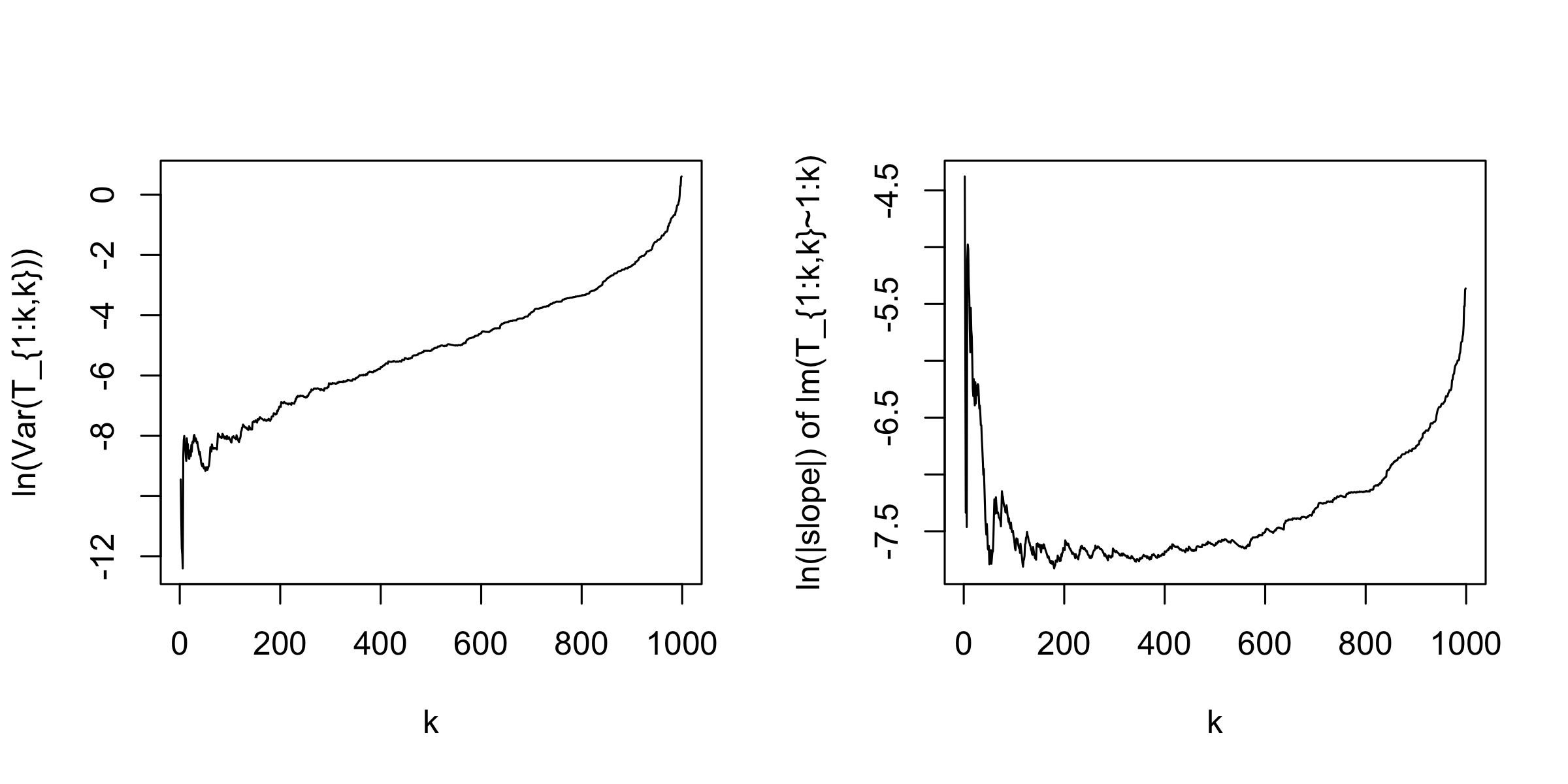}
\caption{Student-t case (10 degrees of freedom). Top left: $T_{b,k}$ for varying lower trimming $b$, for $k=1,51,101,\ldots,1000$.  Top right: Hill plot { (black, solid), together with a mean-of-order $-1$ bias-reduced estimator (blue, dotted)}. Bottom left: empirical variance of the LTH  as a function of $k$. Bottom right: slope of a fitted linear model to the LTH  as a function of $k$.} 
\label{dtht}
\end{figure}

\begin{figure}[hh]
\centering
\includegraphics[width=15cm,trim=.5cm .5cm 0cm .5cm,clip]{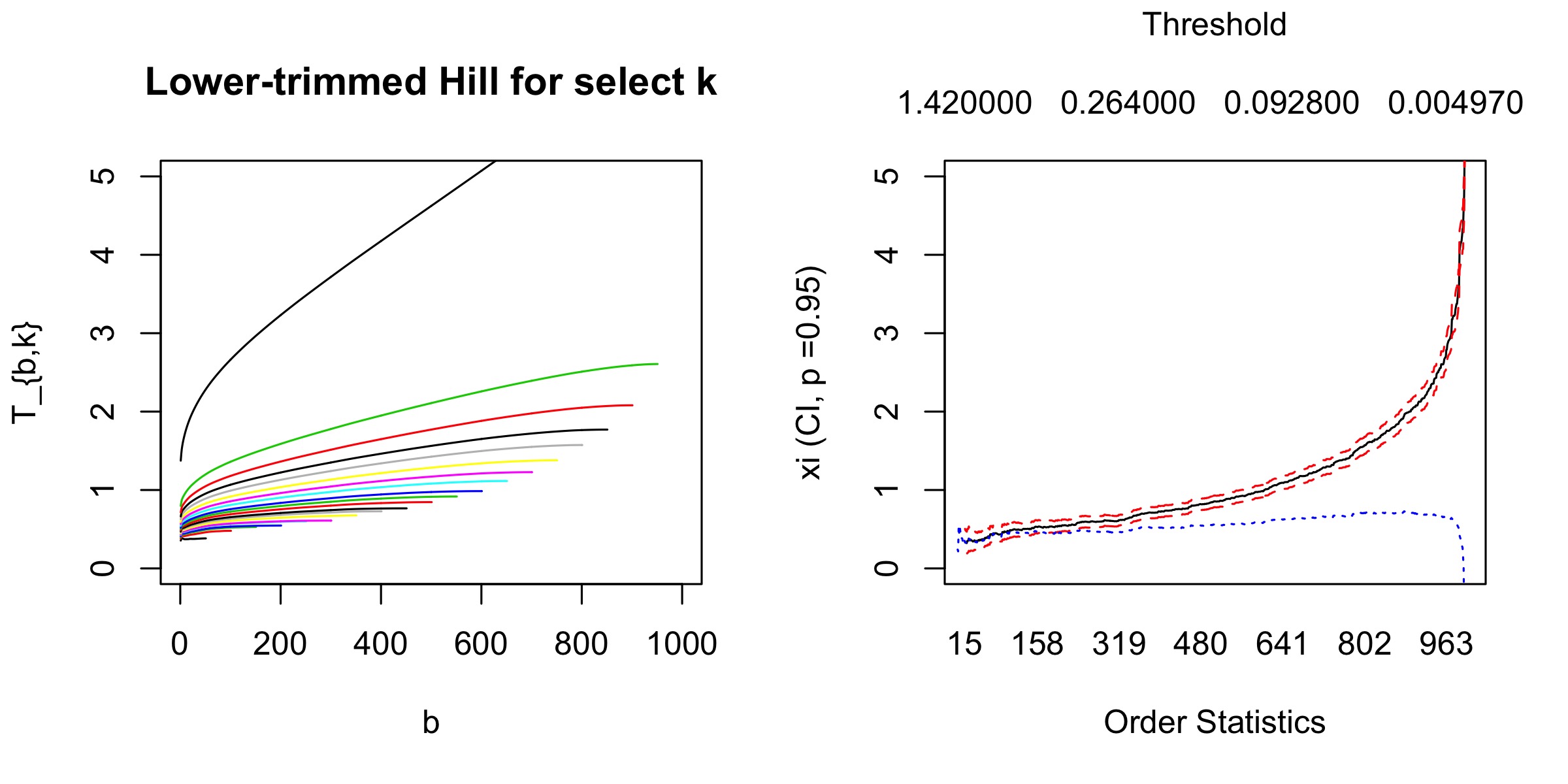}
\includegraphics[width=15cm,trim=.5cm .5cm .5cm 8cm,clip]{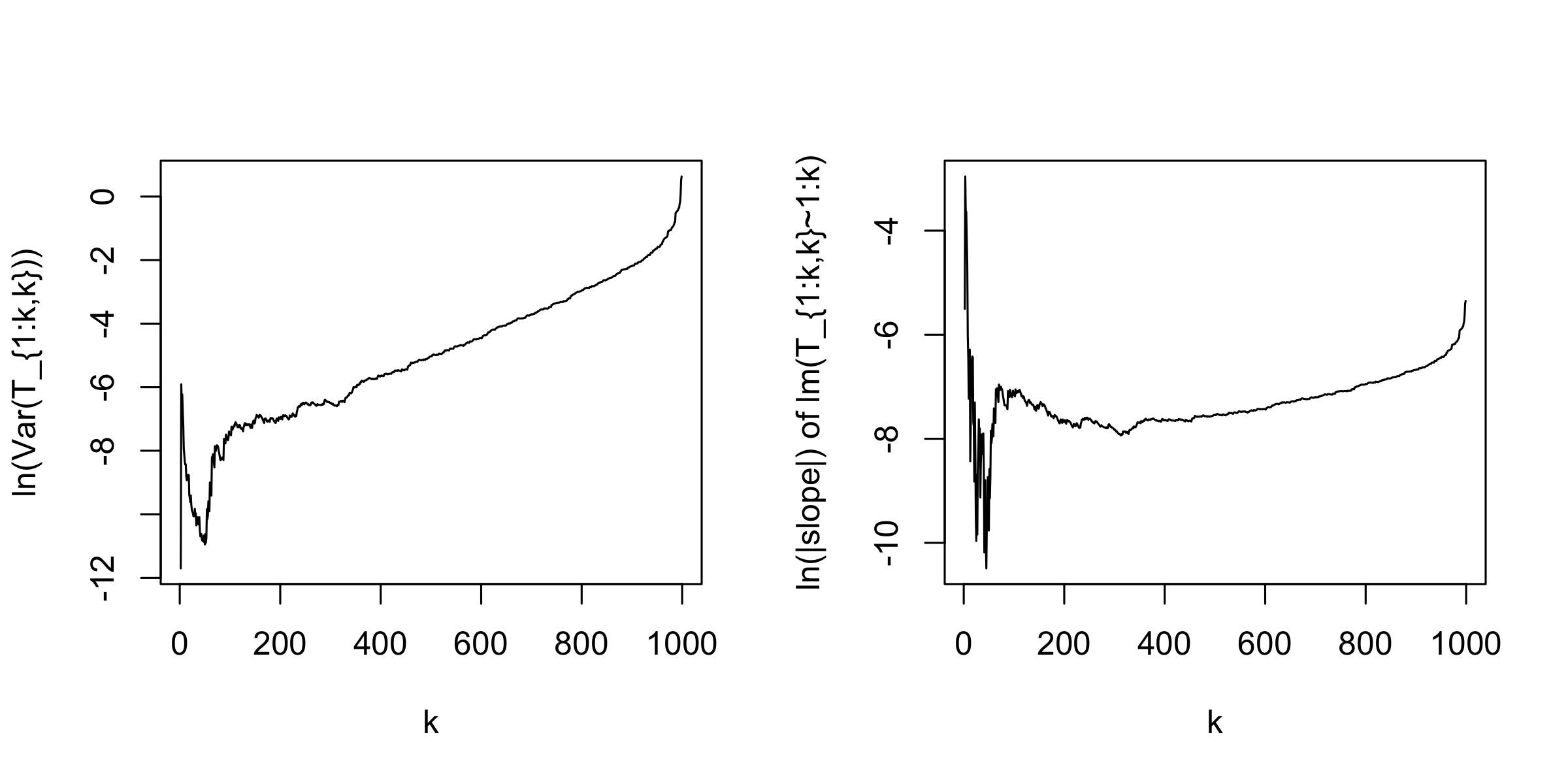}
\caption{Burr case (all parameters set to $1$). Top left: $T_{b,k}$ for varying lower trimming $b$, for $k=1,51,101,\ldots,1000$. Top right: Hill plot { (black, solid), together with a mean-of-order $-1$ bias-reduced estimator (blue, dotted)}. Bottom left: empirical variance of the LTH  as a function of $k$. Bottom right: slope of a fitted linear model to the LTH  as a function of $k$.} 
\label{dthb}
\end{figure}

\begin{figure}[hh]
\centering
\includegraphics[width=15cm,trim=.5cm .5cm 0cm .5cm,clip]{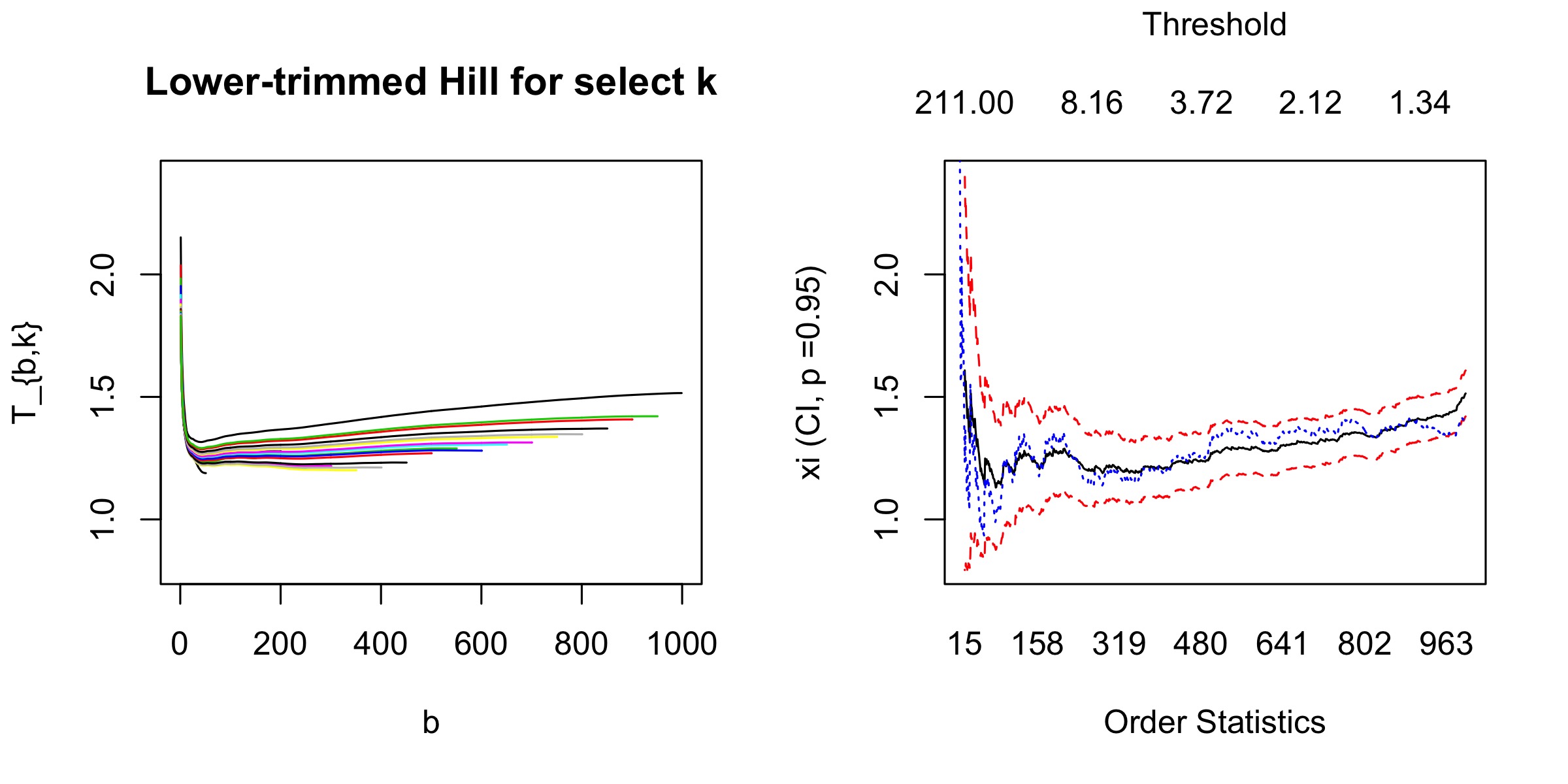}
\includegraphics[width=15cm,trim=.5cm .5cm .5cm 8cm,clip]{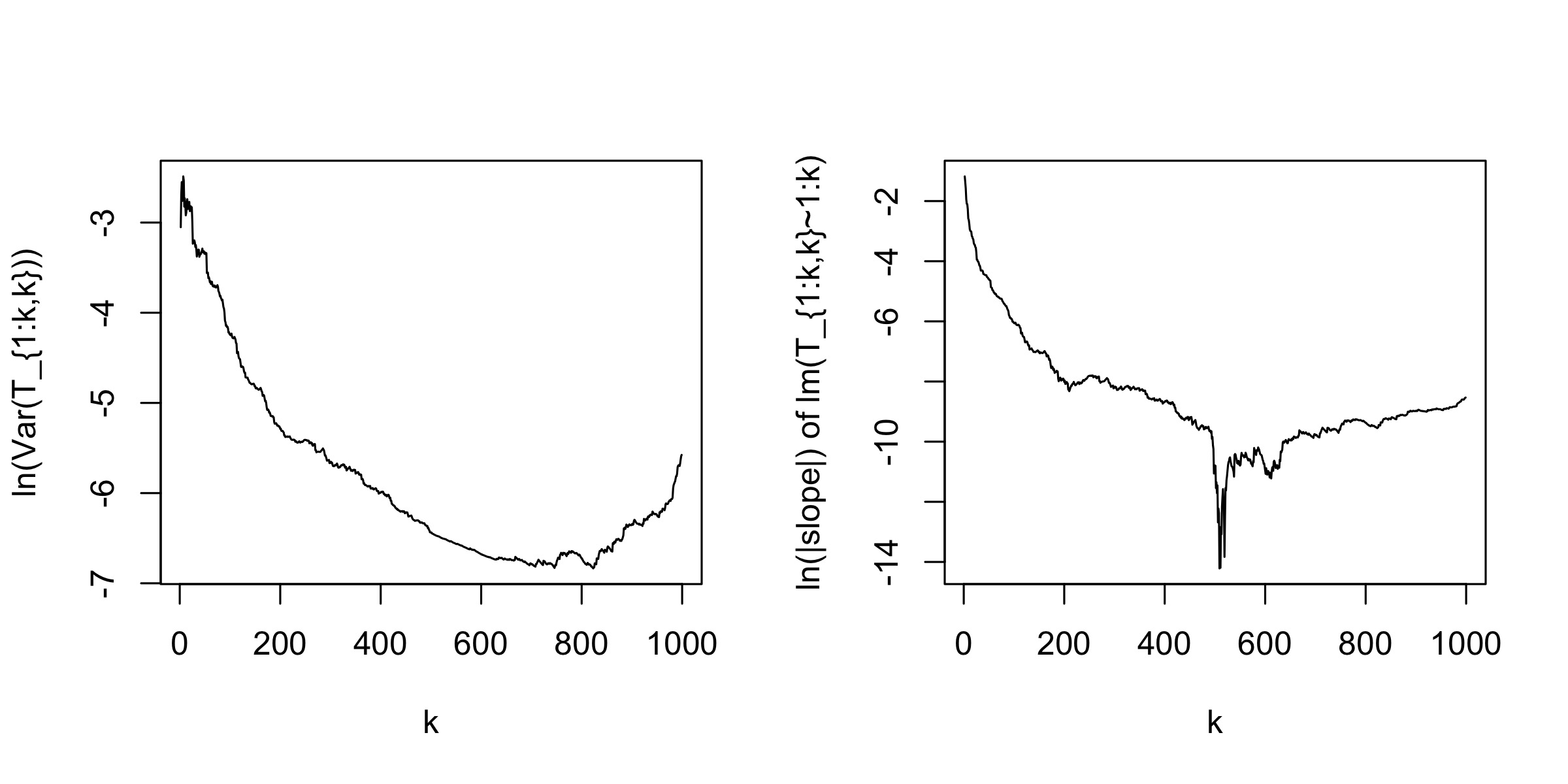}
\caption{Loggamma case (logshape parameter: $3/2$, lograte parameter: $1$). Top left: $T_{b,k}$  for varying lower trimming $b$, for $k=1,51,101,\ldots,1000$. Top right: Hill plot { (black, solid), together with a mean-of-order $-1$ bias-reduced estimator (blue, dotted)}. Bottom left: empirical variance of the LTH as a function of $k$. Bottom right: slope of a fitted linear model to the LTH  as a function of $k$.} 
\label{dthlg}
\end{figure}

 The LTH plots together with usual Hill plots  are shown in the top panels of Figures \ref{dthe}--\ref{dthlg}. { Reduced-bias plots based on Mean-of-Order-$p_0$ estimators are added in Figures \ref{dths2}--\ref{dthlg}. We restrict here to $p_0=-1$ (different choices were also considered, but did not yield substantial differences). } The LTH plots are made for a selection of $k$, from 1 to 1000 by spacings of 50 (1,51,101,...), as a function of the lower trimming $b$. Recall that $b\le k$, so the lines have different domains on the x-axis. Observe that the right end-point of each of the overlaid lines corresponds to the respective point in the Hill plot. The bottom panels of Figures \ref{dthe}--\ref{dthlg} suggest two ways of measuring the aforementioned flatness of the LTH estimator as a function of $b$. The first one computes the empirical variance of $T_{b,k}$, $b=1,\dots,k$, while the second one fits a linear model with independent variable $b=1,\dots,k$ and response variable $T_{b,k}$, and then plots the magnitude of the resulting slope coefficient.\\
 
  For the spliced distribution in Figure \ref{dths2} observe how the LTH estimator becomes horizontal as a function of $b$ when $k$ is close to the (rank of the) splicing point. For smaller $k$, the plot then looks similar to the exact Pareto case. Loosely speaking, the slope of the lines are a  useful visual tool for detecting the number of upper order statistics $k$ after which a Pareto tail is feasible.  
{ The case of the Student-t (Figure  \ref{dtht}) and Burr (Figure \ref{dthb}) distribution show the problem of a large bias for the Hill estimator throughout, where the regime of a Pareto tail is only reached at the most extreme quantiles, and stable $k$ areas are not really available. The Hill plot is roughly a monotone function of the number of order statistics, while  the bias-reduced estimator already departs from the Hill plot at small values of $k$. The $k$ levels indicated by the LTH variance and slope plots confirms that conclusion. Finally, the log-gamma distribution (Figure \ref{dthlg}) with a logarithmic slowly varying function is known to be a difficult case for extreme value analysis: the Hill plot does show a stable area up to $k$ around 500, where the estimates still exhibit a large  bias. The reduced-bias estimates follow the Hill estimates for most of the plot up to $k$ around 500, and this range is also indicated by the LTH slope plot.}\\

 {In the next sections we will develop inferential and data-driven selection and estimation tools on the basis of these graphical tools.}

\section{Regularly varying tails}
\noindent We now move from the simple Pareto sample to a general Fr\'{e}chet domain of attraction, with tails of the form \eqref{Patype}. Denote by $Q$ the quantile function associated to $F$, and define 
\begin{align*}
U(x)=Q(1-1/x), \quad x>1,
\end{align*}
such that the condition \eqref{Patype} is equivalent to
 { 
\begin{align*}
\lim_{t\to\infty}\frac{U(tx)}{U(t)}= x^{\xi}.
\end{align*} 
}
Assumptions on the rate of convergence of the above limit make it possible to obtain explicit results concerning asymptotic properties of the lower-trimmed Hill estimator. Hence, we impose the second order condition
\begin{align}\label{so}
\lim_{t\to\infty}\frac{\log U(tx)-\log U(t)-\xi\log(x)}{Q_0(t)}=\frac{x^{p}-1}{p},
\end{align}
for some regularly varying function $Q_0$ with index $p< 0$.
%
\begin{theorem}\label{theo_tbk_dist}
Under the model \eqref{Patype} and second order condition \eqref{so},  $T_{b,k}$ as defined in \eqref{lth} satisfies the following asymptotic distributional identity, for $n,k,n/k\to \infty$,
\begin{align}\label{astbk}
T_{b,k}\stackrel{d}{=}\xi \frac{\overline E_b+\sum_{j=b+1}^kE_j/j}{1+\sum_{j=b+1}^kj^{-1}}+\frac{Q_0(n/k)}{p}\frac{\frac{((k+1)/b)^p}{1-p}-1}{1+\sum_{j=b+1}^kj^{-1}}(1+o_p(1)),
\end{align}
where $E_1,\dots,E_k$ are i.i.d.\ standard exponential random variables, and  where we use the notation $\overline E_b=b^{-1}\sum_{i=1}^bE_i$.
\end{theorem}

\subsection{Distribution of the average}
Define the average of the $T_{b,k}$ across $b$ as 
\begin{align}\label{barTdef}
\overline T_{k}:=\frac{1}{k} \sum_{b=1}^{k} T_{b,k},
\end{align}
which by Theorem 3.1 satisfies 
\begin{align*}
\overline T_{k}\stackrel{d}{=}\frac{\xi}{k} \sum_{b=1}^{k}\frac{\overline E_b+\sum_{j=b+1}^kE_j/j}{1+\sum_{j=b+1}^kj^{-1}}+\frac{Q_0(n/k)}{pk}\sum_{b=1}^{k}\frac{\frac{((k+1)/b)^p}{1-p}-1}{1+\sum_{j=b+1}^kj^{-1}}(1+o_p(1)).
\end{align*}
We can immediately see that 
\begin{align*}
\E[T_{b,k}]&=\xi +\frac{Q_0(n/k)}{p}\frac{\frac{((k+1)/b)^p}{1-p}-1}{1+\sum_{j=b+1}^kj^{-1}}(1+o_p(1)),\\
\E[\overline T_{k}]&=\xi+\frac{Q_0(n/k)}{pk}\sum_{b=1}^{k}\frac{\frac{((k+1)/b)^p}{1-p}-1}{1+\sum_{j=b+1}^kj^{-1}}(1+o_p(1)),
\end{align*}
so that the asymptotic bias terms can be recognized directly. To ease notation, let us introduce the constants
\begin{align}\label{cbkpconstant}
c_{b,k,p}&:=\frac{1}{p}\cdot \frac{\frac{((k+1)/b)^p}{1-p}-1}{1+\sum_{j=b+1}^kj^{-1}}\approx \frac{1}{p}\cdot \frac{\frac{((k+1)/b)^p}{1-p}-1}{1+\log((k+1)/b)}\\
\overline c_{k,p}&:=\frac{1}{pk}\sum_{b=1}^{k}\frac{\frac{((k+1)/b)^p}{1-p}-1}{1+\sum_{j=b+1}^kj^{-1}}\approx\frac{1}{pk}\sum_{b=1}^{k}\frac{\frac{((k+1)/b)^p}{1-p}-1}{1+\log((k+1)/b)}.\nonumber
\end{align}

\begin{theorem}\label{theo_tk_dist}
The average $\overline T_{k}$ as defined in  \eqref{barTdef}, under model \eqref{Patype} and second order condition \eqref{so} satisfies the following asymptotic distributional identity, for $n,k,n/k\to \infty$,
\begin{align}\label{tkrep}
\overline T_{k}&\stackrel{d}{=}\frac{\xi}{k}\sum_{j=1}^{k}E_j\left[\log(1+\log(k/j))+\frac {ek}{j} \Ei(1+\log(k/j))\right](1+o(1))\\
&\quad+Q_0(n/k) \left[\frac{e^{1-p}}{p(1-p)}\Ei(1-p)-\frac{e}{p}\Ei(1)\right](1+o_p(1)),\nonumber
\end{align}
where $$\Ei(x):=\int_x^\infty e^{-v}/v \dd v,$$ is the exponential integral.
\end{theorem}

Equipped with the representations in terms of exponential variables that we obtained in Theorems  \ref{theo_tbk_dist} and \ref{theo_tk_dist}, we set on to analyze the mean of the empirical variance of $T_{b,k}$ as a function of $b$. 

\begin{theorem}
\color{black}{The mean of the empirical variance of  $\{T_{b,k}; \; 1 \leq b \leq k\}$}, under model \eqref{Patype} and second order condition \eqref{so} satisfies the following asymptotic identity, for $n,k,n/k\to \infty$,
\begin{align*}
&\E\left[\frac{1}{k}\sum_{b=1}^{k}(T_{b,k}-\overline T_k)^2\right]=\frac{C}{k} \xi^2(1+o(1))+Q_0^2(n/k)f(p)(1+o_p(1))\,
\end{align*}
where $C=0.502727$ and
\begin{align}
f(p)&:=\frac{1-e^{1-2p}(1-2p)\Ei(1-2p)-e^{2-2p}\Ei^2(1-p)}{p^2(1-p)^2}\nonumber\\
&\quad+2\frac{e^{2-p}\Ei(1-p)\Ei(1)-1+e^{1-p}(1-p)\Ei(1-p)}{p^2(1-p)}\nonumber\\
&\quad +\frac{1-e\Ei(1)-e^2\Ei^2(1)}{p^2}>0.\label{fp}
\end{align}
\begin{align*}
\end{align*}
\end{theorem}

\begin{figure}[hh]
\centering
\includegraphics[width=8cm,trim=.5cm .5cm .5cm .5cm,clip]{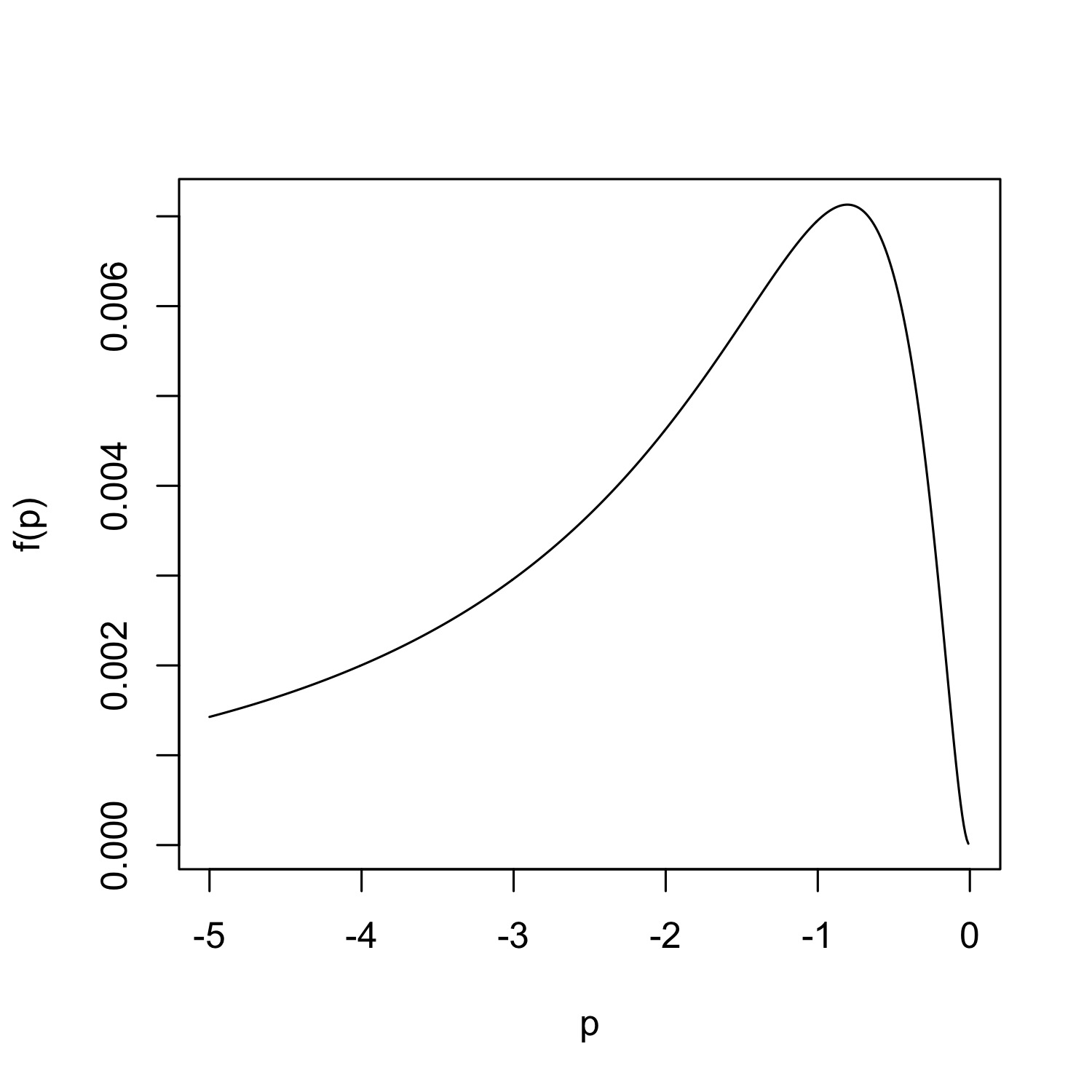}
\caption{Plot of $f(p)$.} 
\label{f_function}
\end{figure}

{  \noindent Notice the fact that $C$ and $f$ are universal. A plot of $f$ as a function of $p$ is given in Figure \ref{f_function}.}

\subsection{Optimal $k$ in the Hall class}

\noindent We now make a further assumption on the regularly varying class, in order to get an explicit form of $Q_0$. Concretely, we assume the Hall class (\cite{hall82}), which satisfies the property
\begin{align}\label{hallclassquantiles}
U(x)=Ax^{\xi}(1+Dx^{p}(1+o(1))), \quad x\to\infty.
\end{align}
An immediate consequence then is the  explicit expression
 {
\begin{align*}
Q_0(x)=pDx^{p}(1+o(1)).
\end{align*}
}
Hence,
\begin{align}\label{expempvar}
&\E\left[\frac{1}{k}\sum_{b=1}^{k}(T_{b,k}-\overline T_k)^2\right]=\frac{C}{k} \xi^2(1+o(1))+Q_0^2({n\over k})f(p)(1+o_p(1)).
\end{align}
Recall that the classical Hill estimator for this class has AMSE given by
\begin{align*}
\frac{\xi^2}{k}+\left(\frac{ Q_0(n/k)}{1-p}\right)^2,
\end{align*}
which is minimized for
\begin{align}
k_0^\ast &\sim (Q^2_0(n))^{-1/(1-2p)}\left(\frac{\xi^2(1-p)^2}{-2p}\right)^{1/(1-2p)}\nonumber\\
&=\left(\frac{n^{-2p}\xi^2(1-p)^2}{-2p^3D^2}\right)^{1/(1-2p)},\label{amsemink}
\end{align}
see e.g.\ \cite[p.125]{beirlant2006}). In a similar way, the minimizer of \eqref{expempvar} is simply
\begin{align}\label{kast}
k^\ast &\sim (Q^2_0(n))^{-1/(1-2p)}\left(\frac{C\xi^2}{-2pf(p)}\right)^{1/(1-2p)}\nonumber\\
&=\left(\frac{n^{-2p}C\xi^2}{-2p^3D^2f(p)}\right)^{1/(1-2p)}.
\end{align}
\color{black}{Hence from \eqref{amsemink} and \eqref{kast} we obtain a simple expression of the optimal threshold $k_0^\ast$ of the Hill estimator in terms of $k^\ast$:
\begin{equation}
k_0^\ast = k^\ast \left(\frac{C}{(1-p)^2f(p)}\right)^{-1/(1-2p)}.
\label{kastkast}
\end{equation}
}

\subsection{Interpretation of $\overline{T}_k$ as a weighted Hill estimator}\label{secw}
\begin{figure}[hh]
\centering
\includegraphics[width=7cm,trim=0cm 0cm 0cm 0cm,clip]{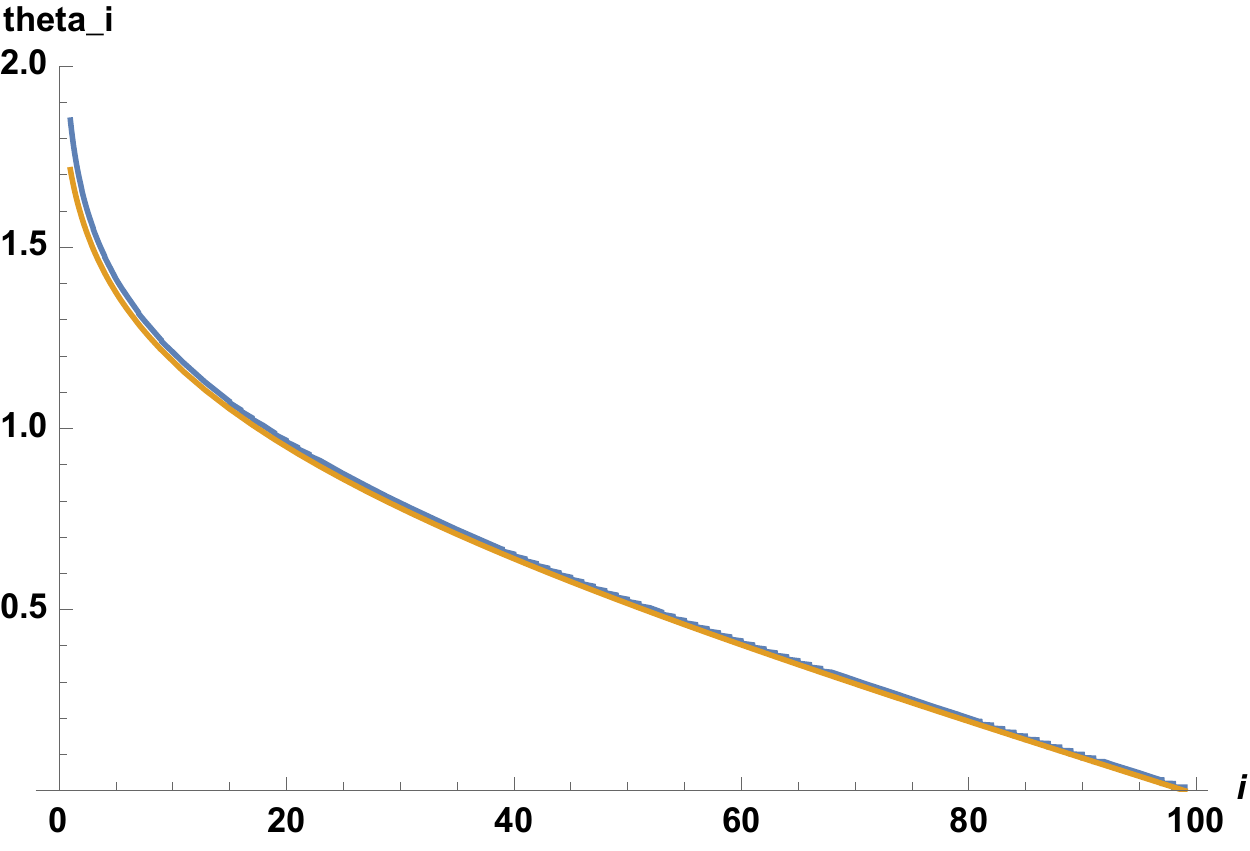}
\includegraphics[width=7cm,trim=0cm 0cm 0cm 0cm,clip]{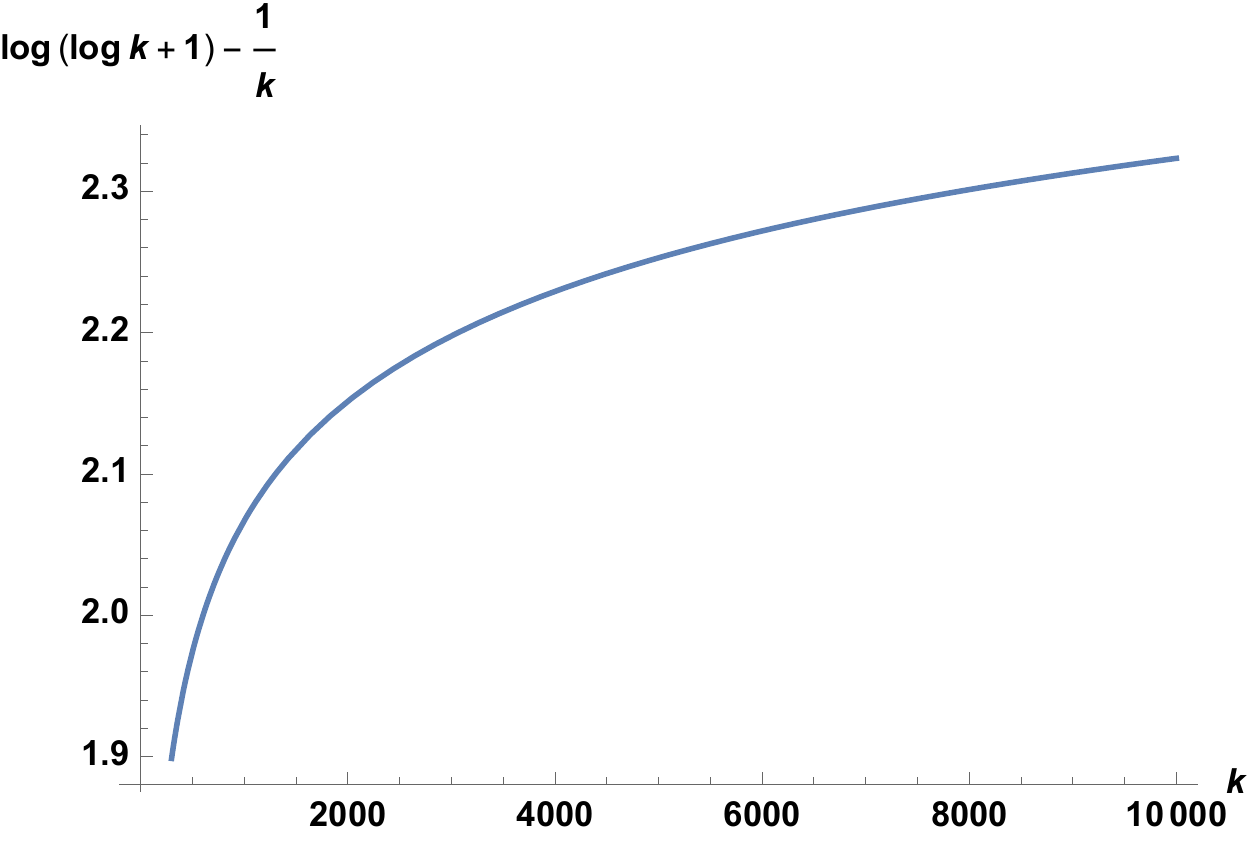}
\caption{Left panel: for $k=100$, the true $\theta_i$ (blue) and the asymptotic approximation { from equation} \eqref{asympapproxtheta} (orange) as a function of $i$. Right panel: Leading terms of the series expansion of $\theta_1$ with respect to $k$.} 
\label{asymps}
\end{figure}

Observe that, for fixed $k$, 
\begin{align}\label{hillwiththeta}
\overline T_{k}=\frac{1}{{k}}\sum_{b=1}^{{k}}T_{b,{k}}=\frac{1}{k}\sum_{i=1}^{k}\theta_i \log(Y_{i,k}),
\end{align}
with
\begin{align*}
\theta_i:=\sum_{b=i}^{k}\frac{1}{b(1+\sum_{j=b+1}^{k} j^{-1})},
\end{align*}
so that one can interpret the estimator $\overline T_{k}$ as a modification of the classical Hill estimator that uses different weights for different order statistics. It is not hard to see that asymptotically the correction factors behave like
\begin{align}\label{asympapproxtheta}
\theta_i\sim \log\left(\frac{\log(i/k)-1}{\log(1-1/k)-1}\right),\;k\to\infty.
\end{align}
Figure \ref{asymps}(left) highlights the accuracy of this approximation for $k=100$ across different values of $i$, and also illustrates the fact that the largest data point receives a weight of almost 2 in this case, whereas on from the 20th-largest observation the weight is lower than for the classical Hill estimator, and the weight diminishes for smaller data points. Note that, as $k$ increases, the weight of the largest observation grows above any bound, but extremely slowly, namely 
 $$\theta_1=\log(\log(k)+1)-1/k+O(1/k^3).$$
Figure \ref{asymps}(right) illustrates that even for a value as large as $k=10000$, $\theta_1$ is still below 2.4. 

\section{A ratio statistic}
Once a $k^\ast$ has been selected, it is important to be able to statistically assess whether the remaining upper tail differs significantly from the one of a pure Pareto. In order to recognize whether a Pareto tail has been achieved or not, we have seen that flatness of the lower-trimmed Hill estimator is desirable. Inspired by the T-statistic introduced in \cite{hilltrim}, we introduce the ratio statistics
\begin{align*}
R_{b,k}=\frac{T_{b+1,k}}{T_{b,k}}, \quad b=1,\dots,k-1,
\end{align*}
quantities which we expect to be close to one. Although these statistics do not have the property of being i.i.d.\ and hence test sizes have to be calibrated using Monte Carlo simulation, an advantage which carries over to the present setting is that they do not depend on $\xi$. Indeed, { defining $\omega(b,k)=b(1+\sum_{j=b+1}^kj^{-1})$, we have}
\begin{align*}
R_{b,k}\stackrel{d}{=}\frac{\omega(b,k)}{\omega(b+1,k)}\left(1+\frac{\log\left(\Gamma_{b+1}/\Gamma_{k+1}\right)}{\sum_{i=1}^b\log\left(\Gamma_{i}/\Gamma_{k+1}\right)}\right),
\end{align*}
by the order statistics property of the Poisson process, where, $\Gamma_m=\sum_{i=1}^m E_i$, and $E_i$, $i=1,2,\dots$, is an i.i.d.\ sequence of independent unit-rate exponential random variables. This  invariance with respect to the $\xi$ parameter permits to assess the goodness of selection of a threshold $k^\ast$ as follows:

\begin{enumerate}
\item Simulate the $R_{b,k^\ast}$ statistics $N_{MC}$ times, and call them
\begin{align*}
R^m_{b,k^\ast}, \quad m=1,\dots, N_{MC}, \: b=2,\dots k^\ast-1.
\end{align*}
\item For fixed $\alpha\in (0,1)$, find the empirical $\alpha/2$ and $1-\alpha/2$ quantiles corresponding to each of the $b=2,\dots,k^\ast-1$ samples,
\begin{align*}
R^m_{b,k^\ast}, \quad m=1,\dots, N_{MC},
\end{align*}
and call them $(q_1,q_2)_2,\dots, (q_1,q_2)_{k^\ast-1}$.
\item Count the proportion of the $N_{MC}$ trajectories 
\begin{align*}
{ (R^m_{b,k^\ast})_{ b=2,\dots, k^\ast-1}},
\end{align*}
which for { \textit{some}} $b\in \{2\dots k^\ast-1\}$ fall outside of their confidence interval $(q_1,q_2)_b$. Call this proportion $\alpha_r$.
\item {  The proportion $\alpha_r$ is the global level of the test, and the algorithm stops when it is close enough to a pre-specified level (typically $0.05$). If $\alpha_r$ is larger (smaller) than the pre-specified level, go to Step (2) and decrease (increase) $\alpha$.}
\item Plot the $R_{b,k^\ast}$, $b=2,\dots, k^\ast-1$, from the data, together with the last set of quantiles $(q_1,q_2)_1,\dots, (q_1,q_2)_{k^\ast{ -1}}$. It is also a good idea, for visualization, to plot the standardized version
\begin{align*}
\frac{R_{b,k^\ast}-q_{1,b}}{q_{2,b}-q_{1,b}},\quad b=2,\dots,k^\ast-1,
\end{align*}
which for a pure Pareto tail is expected by construction to lie (as a trajectory) between $0$ and $1$ in $100(1-\alpha_r)\%$ of the cases. {  Here,  we have used the notation $(q_1,q_2)_b=(q_{1,b},q_{2,b})$.}

\end{enumerate}

\begin{figure}[hh]
\centering
\includegraphics[width=16cm,trim=.5cm 2cm .5cm 8cm,clip]{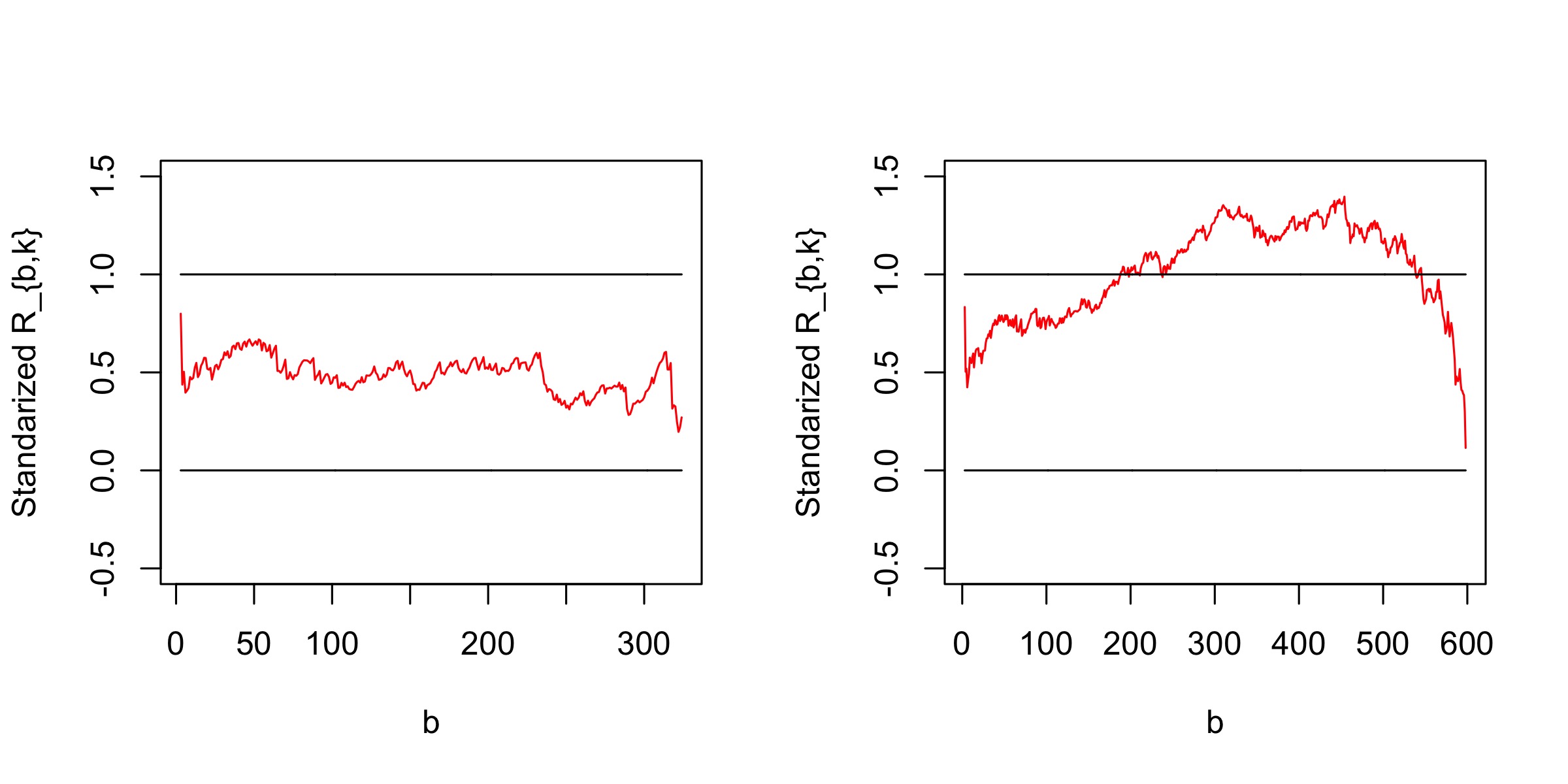}
\caption{Standarized R-statistic for the Burr sample of Figure \ref{dthb} (all parameters set to $1$), for two choices of threshold: $k=326,600$, respectively. $N_{MC}=10000$ and $\alpha=0.05$.} 
\label{tstatburr}
\end{figure}

\begin{example}\normalfont
For the Burr sample of Figure \ref{dthb}, we compare taking $k^\ast=326$ and $k^\ast=600$ in the plots of Figure \ref{tstatburr}. The first number, $k^\ast=326$ is precisely the one that minimizes the expected empirical variance, according to the parameters of the Burr sample and to formula \eqref{kast}, with $p$ chosen to be $-1$. The number of Monte Carlo simulations was in each case $N_{MC}=10000$, and the significance level is $\alpha=0.05$. Observe how the fit is good for $k=326$,  but is outside the bands for $k=600$.
\end{example}

\begin{remark}\normalfont
This approach can only be considered as a selection procedure itself if the corresponding sequential testing is adjusted to have the correct size. In other words, if the above algorithm is used multiple times to choose $k$, the rejection probability will exceed the desired $\alpha$ level. An alternative is to take sequential values of $k$ into the algorithm, which makes the routine highly computationally intensive. Hence, we presently recommend it solely as a goodness of selection evaluation. \end{remark}

\section{Simulations}\label{secsim}
We perform a simulation study based on three different and common distributions which belong to the Hall class \eqref{hallclassquantiles}. We consider simulating $N_{{sim}}=1000$ times from the following three distributions, with four sub-cases for each distribution, for varying sample size and parameters:
\begin{itemize}
\item The Burr distribution, with tail given by
\begin{align*}
\overline F(x)=\left(\frac{\eta}{\eta+x^\tau}\right)^\lambda, \;x>0,\quad \eta,\tau,\lambda>0,
\end{align*}
which implies by Taylor expansion that
\begin{align*}
\xi=\frac{1}{\lambda\tau},\quad A=\eta^{1/\tau},\quad D=-\frac{1}{\tau},\quad p=-\frac{1}{\lambda}.
\end{align*}
We consider for $n=100,\,500$ the two sets of parameters $\eta=1$, $\lambda=2$, $\tau=1/2$; and $\eta=3/2$, $\lambda=1/2$, $\tau=2$.\\

\item The Fr\'{e}chet distribution with tail 
\begin{align*}
\overline F(x)=1-\exp(-x^{-\alpha}), \quad \alpha>0,
\end{align*}
which implies
\begin{align*}
\xi=\frac{1}{\alpha},\quad A=1,\quad D=-\frac{1}{2\alpha},\quad p=-1.
\end{align*}
We consider for $n=100,\,500$ the two parameters $\alpha=1,\,1/2$.\\

\item The Generalized Pareto Distribution (GPD) distribution, with tail given by
\begin{align*}
\overline F(x)=\left(1+\frac{\gamma x}{\sigma}\right)^{-1/\gamma}, \quad \gamma,\sigma>0,
\end{align*}
which implies
\begin{align*}
\xi=\gamma,\quad A=\frac{\sigma}{\gamma},\quad D=-1,\quad p=-\gamma.
\end{align*}
We consider for $n=100,\,500$ the two sets of parameters $\gamma=1/2$, $\sigma=2$; and $\gamma=5/2$, $\sigma=1$.\\

{ \item The Student-t distribution with $m$ degrees of freedom. The tail is given in terms of  hypergeometric and Gamma functions. Since this distribution is symmetric around zero, we take the absolute values of the data, which preserves the tail behaviour. We have that 
\begin{align*}
\xi=\frac{1}{m},\quad p=-\frac{2}{m}.
\end{align*}
We consider, for $n=100,\,500$, the two sets of parameters $m=2,\, 10$.}\\
\end{itemize}

\noindent  For each sample we evaluate the Hill estimator $$H_k=T_{k,k}$$ and the averaged trimmed estimator  $$\overline{T}_k=\frac{1}{{k}}\sum_{b=1}^{{k}}T_{b,{k}}$$ at three particular choices of $k$. {\color{black} Note that these \color{black}{threshold} choices are designed for the Hill estimator, but will \color{black}{turn out} sensible for the latter estimator as well.}\\

\begin{itemize}
\item[(i)] {\color{black}We use the popular procedure of \cite{guillou2001diagnostic} as a benchmark for finding the optimal choice of $k$, and denote the resulting tail estimators by $H_{\hat k_{GH}}$, $\overline T_{\hat k_{GH}}$. Such \color{black}{a} threshold selection procedure has been subject to comparisons (\color{black}{both} in \cite{guillou2001diagnostic} itself and in \cite{beirlant2002exponential}) to other alternatives like \cite{danielsson2001using} and \cite{drees1998selecting}.  {We also refer to \cite{Schneider} for a recent paper which was developed independently around the same time as the present article.}}\\
\item[(ii)]  { An estimator $\hat{k}_0^\ast$ of $k_0^\ast$ from \eqref{amsemink} is obtained as follows. Motivated by \eqref{expempvar}, we compute $\hat{k}^\ast$ as the minimizer of the empirical variance (the search beginning at 1/5 of the sample size, to avoid degeneracies) of the trimmed Hill estimator, as a function of $b$, and using \eqref{kastkast} to set
\begin{align*}
\hat{k}^*_0:=\hat{k}^\ast\left(\frac{C}{(1-p)^2f(p)}\right)^{-1/(1-2p)}.
\end{align*}
}
Observe that while we still have to input $p$, here prior knowledge of $\xi,D$ is no longer needed. We choose $p=-1$ as the canonical choice.\\ 
\item[(iii)] As in (ii), but using the true parameter of $p$, in order to quantify how the removal of a potential misspecification of $p$ by the canonical choice $p=-1$ affects the estimators.  {The results obtained when using an estimator of $p$, such as given by \cite{Alves}, rather than fixing it at a fixed value, 
indicated that the 	additional variability by including an additional estimator does not lead to further improvements at smaller sample sizes. This confirms the observations made for instance in \cite{drees1998selecting} and \cite{beirlant2002exponential}.} \\
\end{itemize}
We then plot the bias, variance and MSE of each resulting estimator as a function of $k$. \\

\begin{figure}[hh]
	\centering
	\includegraphics[width=11cm,trim=.5cm 0cm 1.5cm .5cm,clip]{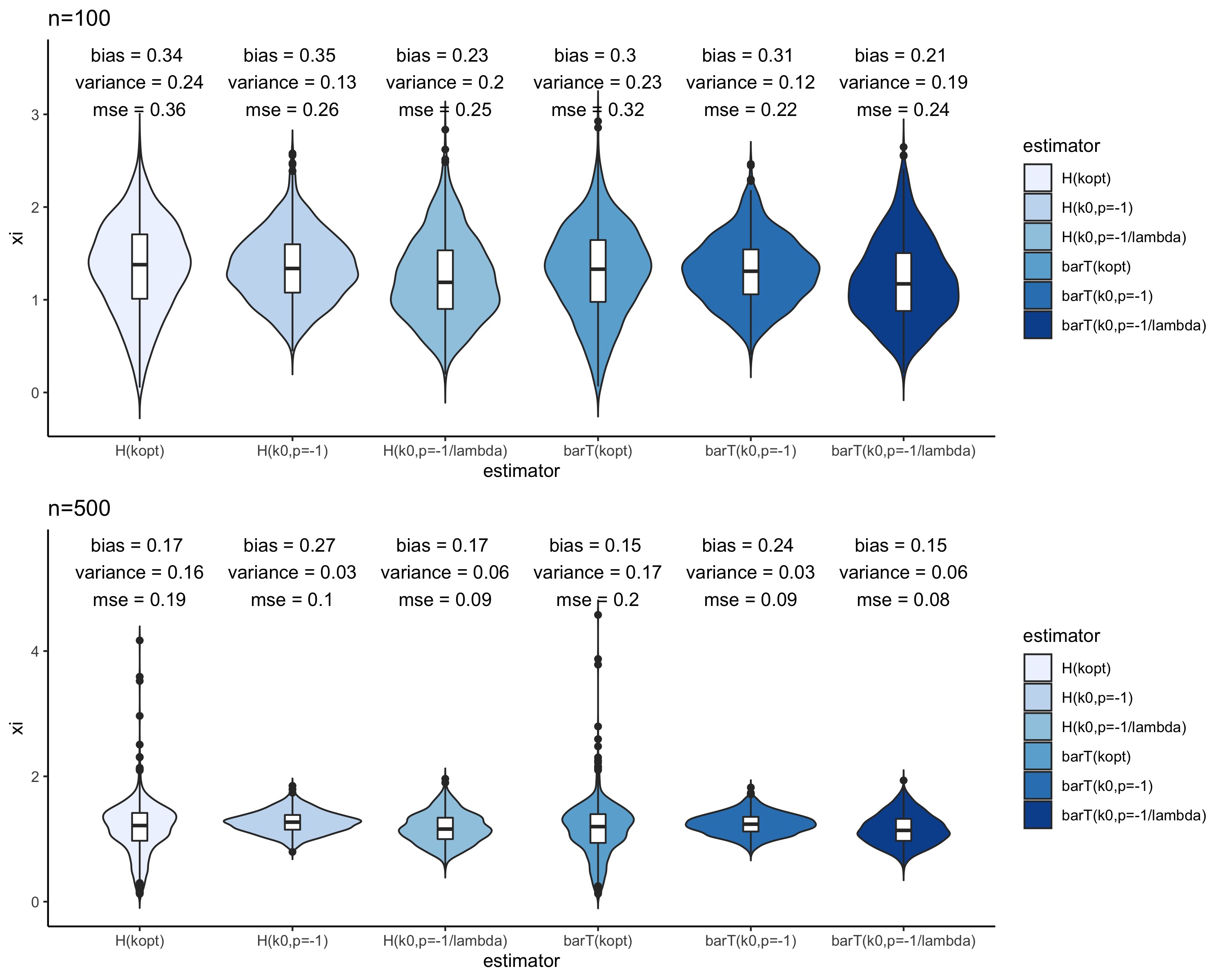}
	\includegraphics[width=12cm,trim=.5cm .5cm .5cm 0cm,clip]{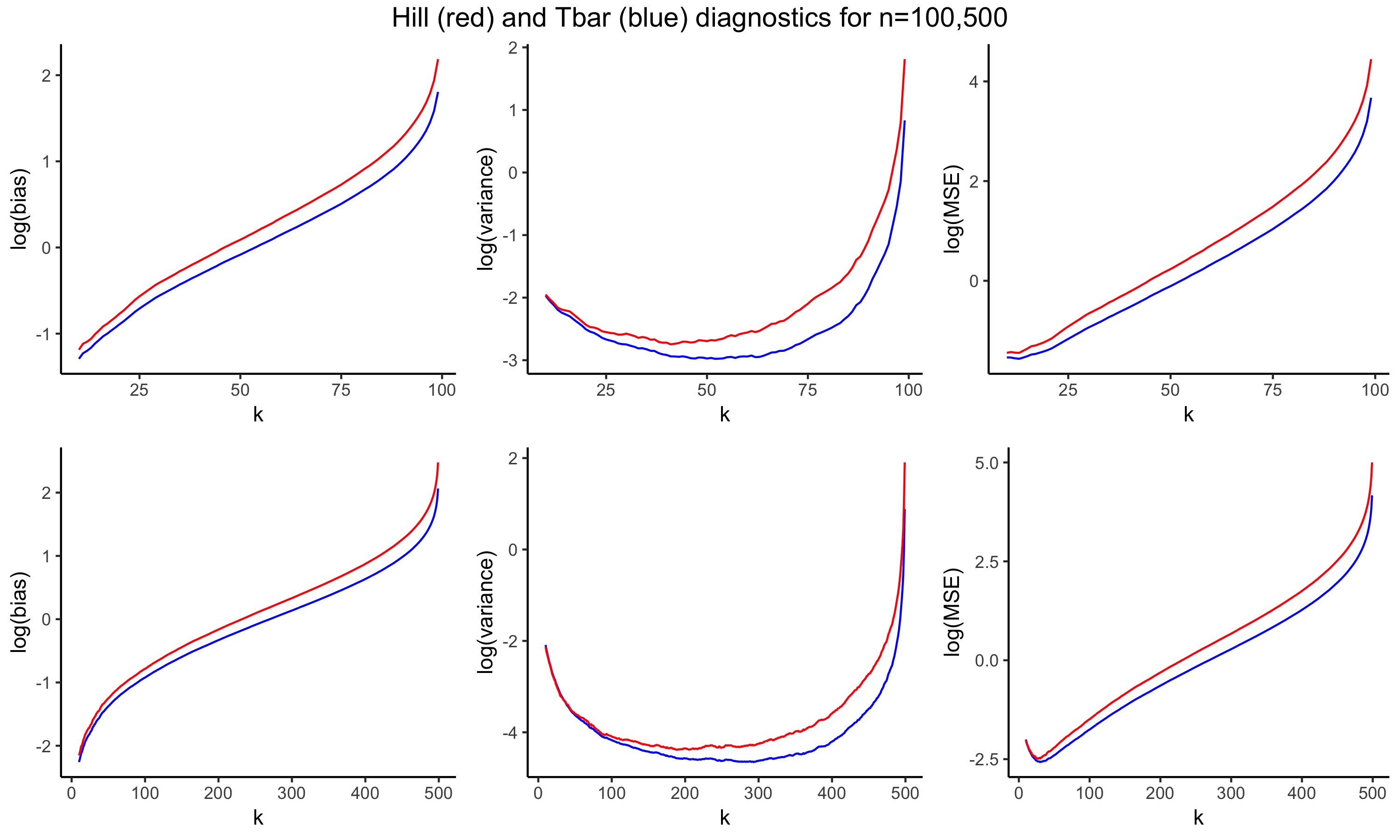}
	\caption{Burr distribution, parameters $\eta=1$, $\lambda=2$, $\tau=1/2$. Top: Violin plots for $n=100,500$ of the estimators $H_{\hat k_{GH}}$, $H_{\widehat k^\ast_0,\,p=-1}$, $H_{\widehat k^\ast_0,\,p=-1/\lambda}$, $\overline T_{\hat k_{GH}}$, $\overline T_{\widehat k^\ast_0,\,p=-1}$, $\overline T_{\widehat k^\ast_0,\,p=-1/\lambda}$. Bottom: diagnostics of $\overline T_k$ (blue) and $H_k$ (red) as a function of $k$.} 
	\label{burr_n100_n500_params1}
\end{figure}

\begin{figure}[hh]
\centering
\includegraphics[width=11cm,trim=.5cm 0cm 1.5cm .5cm,clip]{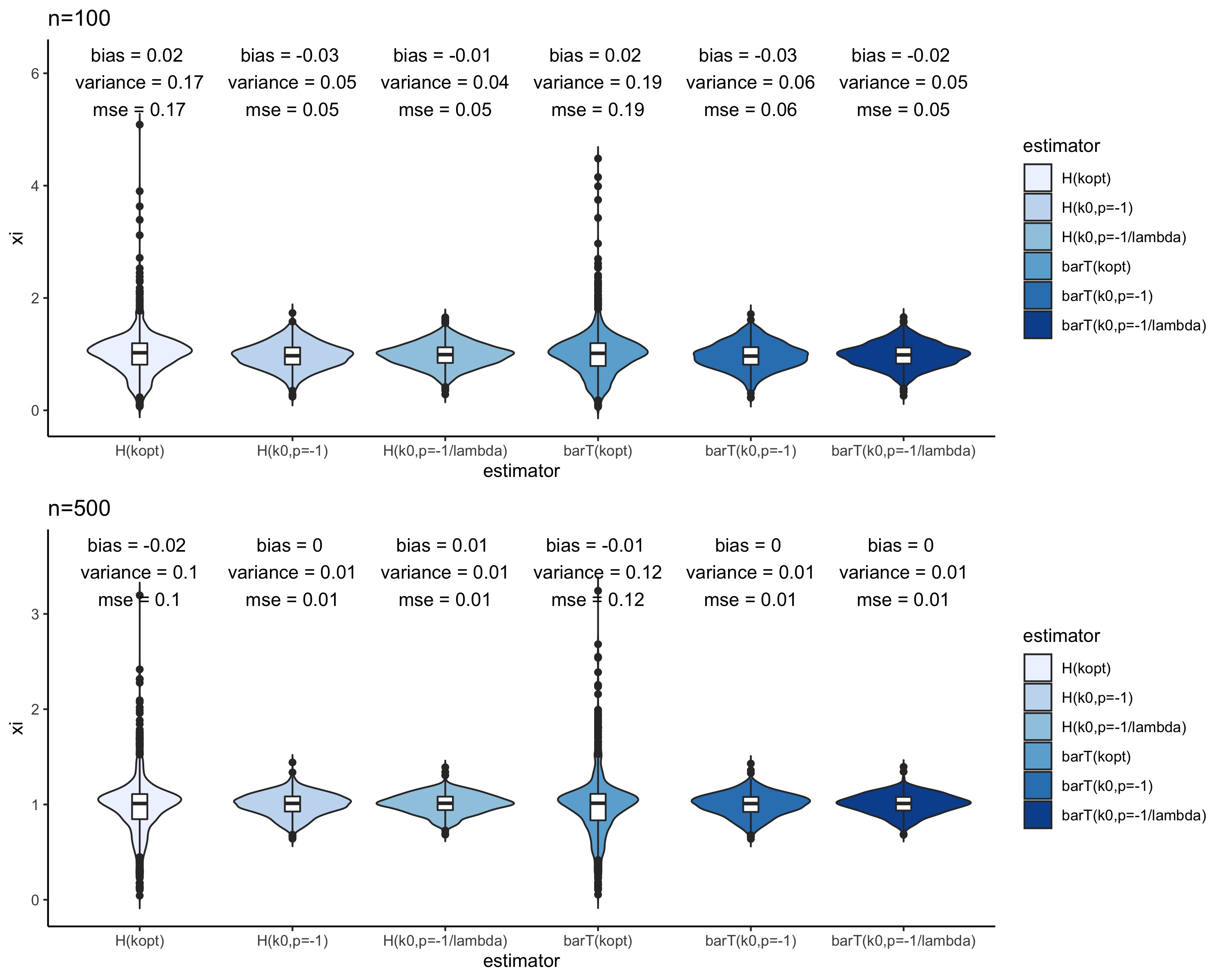}
\includegraphics[width=12cm,trim=.5cm .5cm .5cm 0cm,clip]{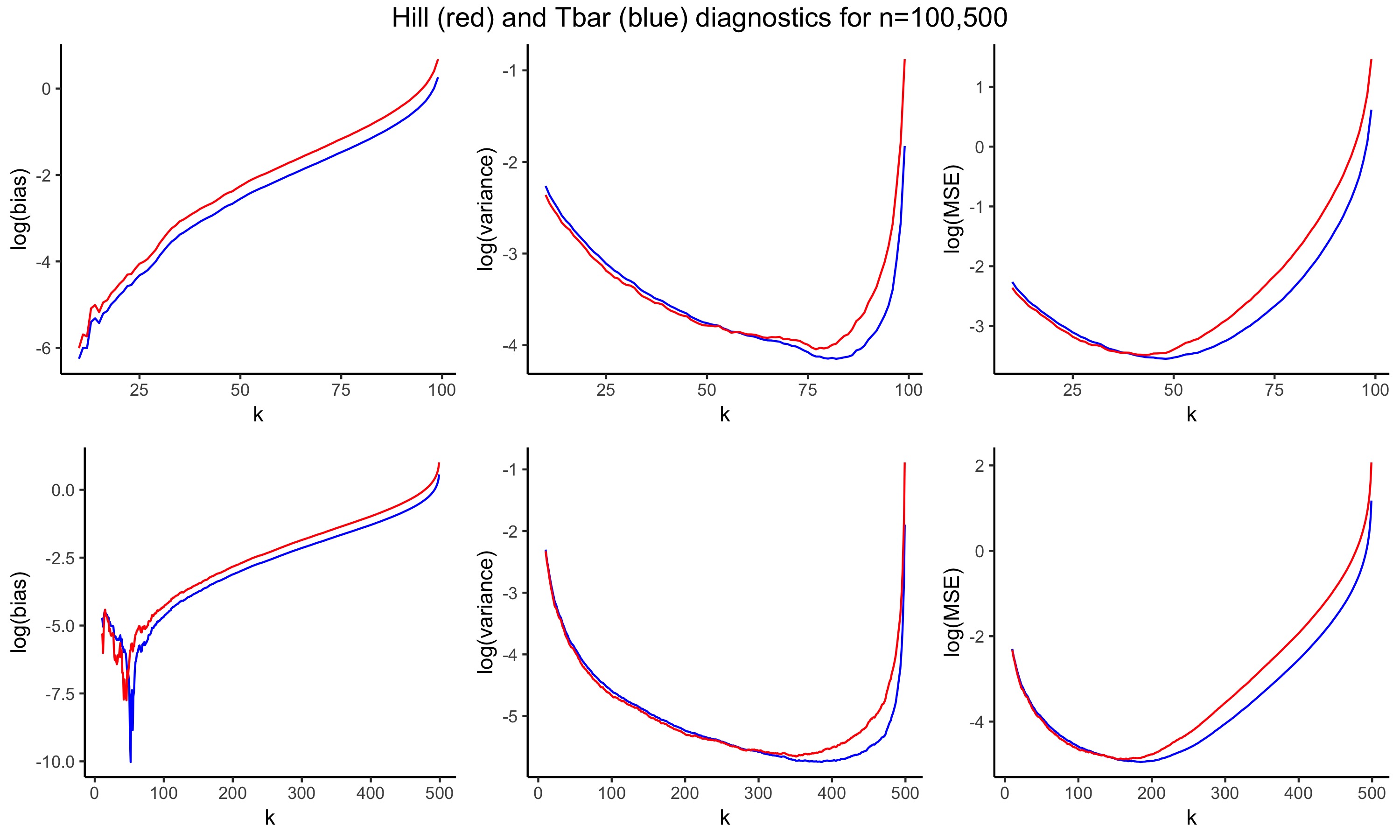}
\caption{Burr distribution, parameters $\eta=3/2$, $\lambda=1/2$, $\tau=2$. Top: Violin plots for $n=100,500$ of the estimators $H_{\hat k_{GH}}$, $H_{\widehat k^\ast_0,\,p=-1}$, $H_{\widehat k^\ast_0,\,p=-1/\lambda}$, $\overline T_{\hat k_{GH}}$, $\overline T_{\widehat k^\ast_0,\,p=-1}$, $\overline T_{\widehat k^\ast_0,\,p=-1/\lambda}$. Bottom: diagnostics of $\overline T_k$ (blue) and $H_k$ (red) as a function of $k$.} 
\label{burr_n100_n500_params2}
\end{figure}

\begin{figure}[hh]
\centering
\includegraphics[width=10cm,trim=.5cm 0cm 1.5cm .5cm,clip]{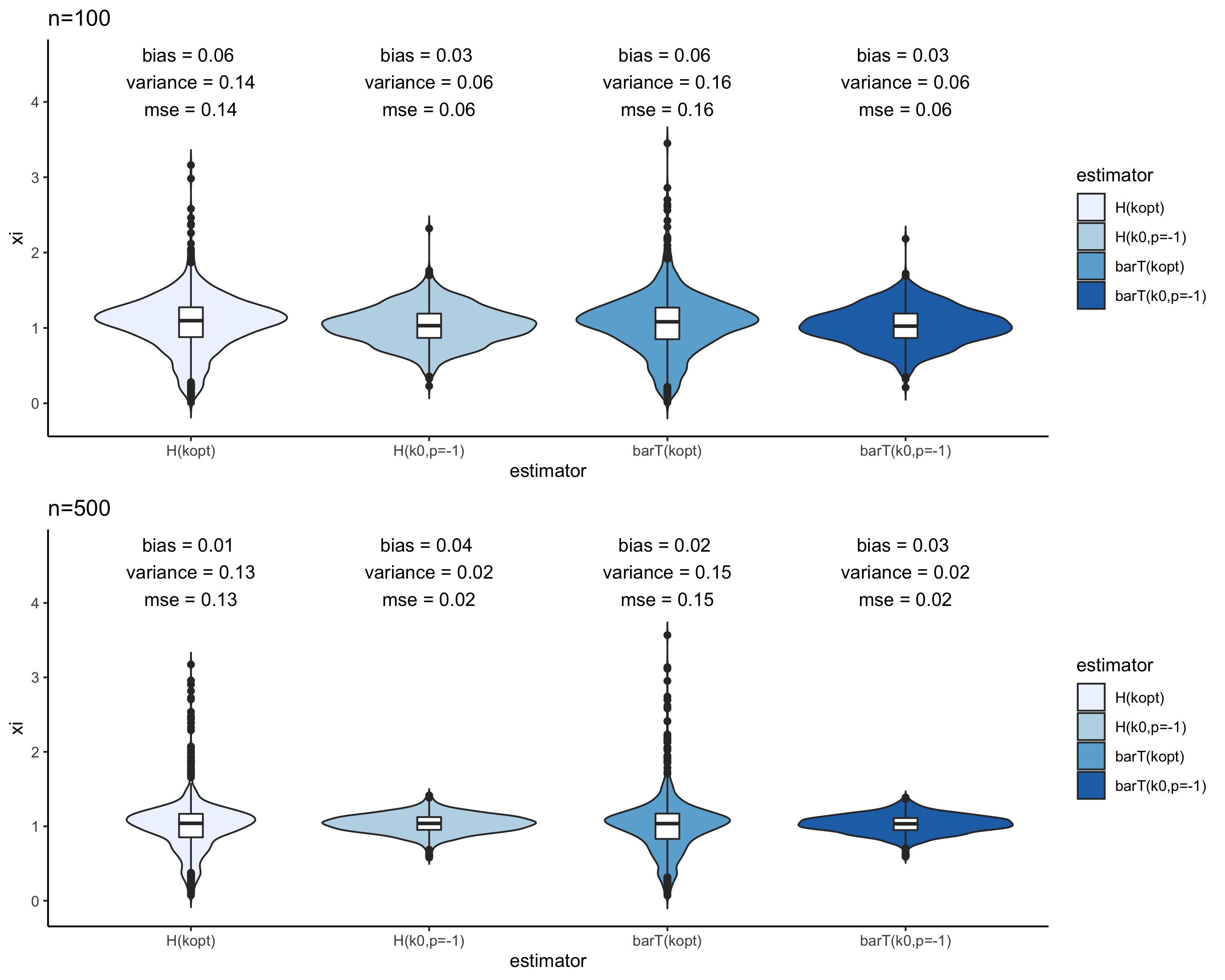}
\includegraphics[width=12cm,trim=.5cm .5cm .5cm 0cm,clip]{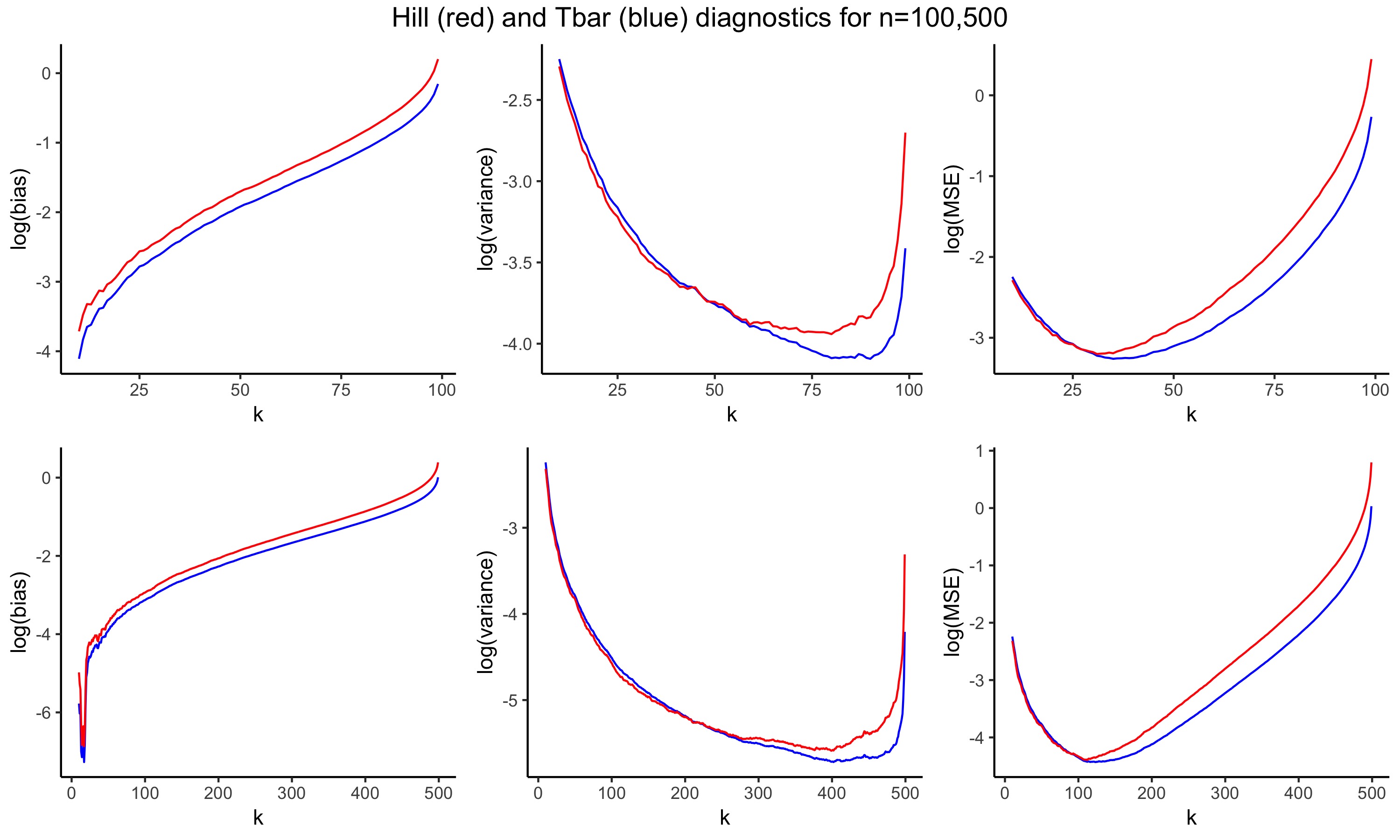}
\caption{Fr\'{e}chet distribution, parameter $\alpha=1$. Top: Violin plots for $n=100,500$ of the estimators $H_{\hat k_{GH}}$, $H_{\widehat k^\ast_0,\,p=-1}$, $\overline T_{\hat k_{GH}}$, $\overline T_{\widehat k^\ast_0,\,p=-1}$. Bottom: diagnostics of $\overline T_k$ (blue) and $H_k$ (red) as a function of $k$.} 
\label{frechet_n100_n500_params1}
\end{figure}

\begin{figure}[hh]
\centering
\includegraphics[width=10cm,trim=.5cm 0cm 1.5cm .5cm,clip]{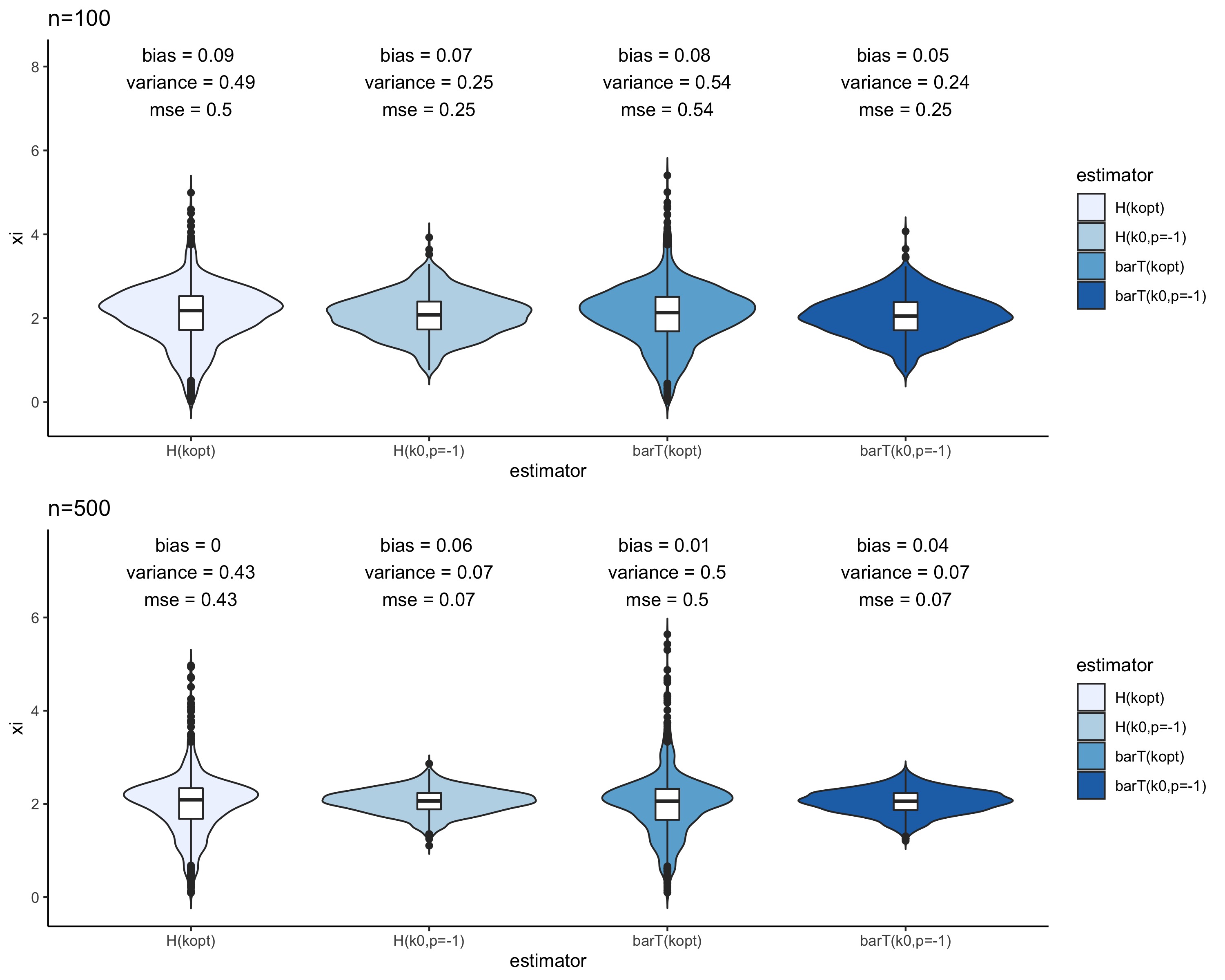}
\includegraphics[width=12cm,trim=.5cm .5cm .5cm 0cm,clip]{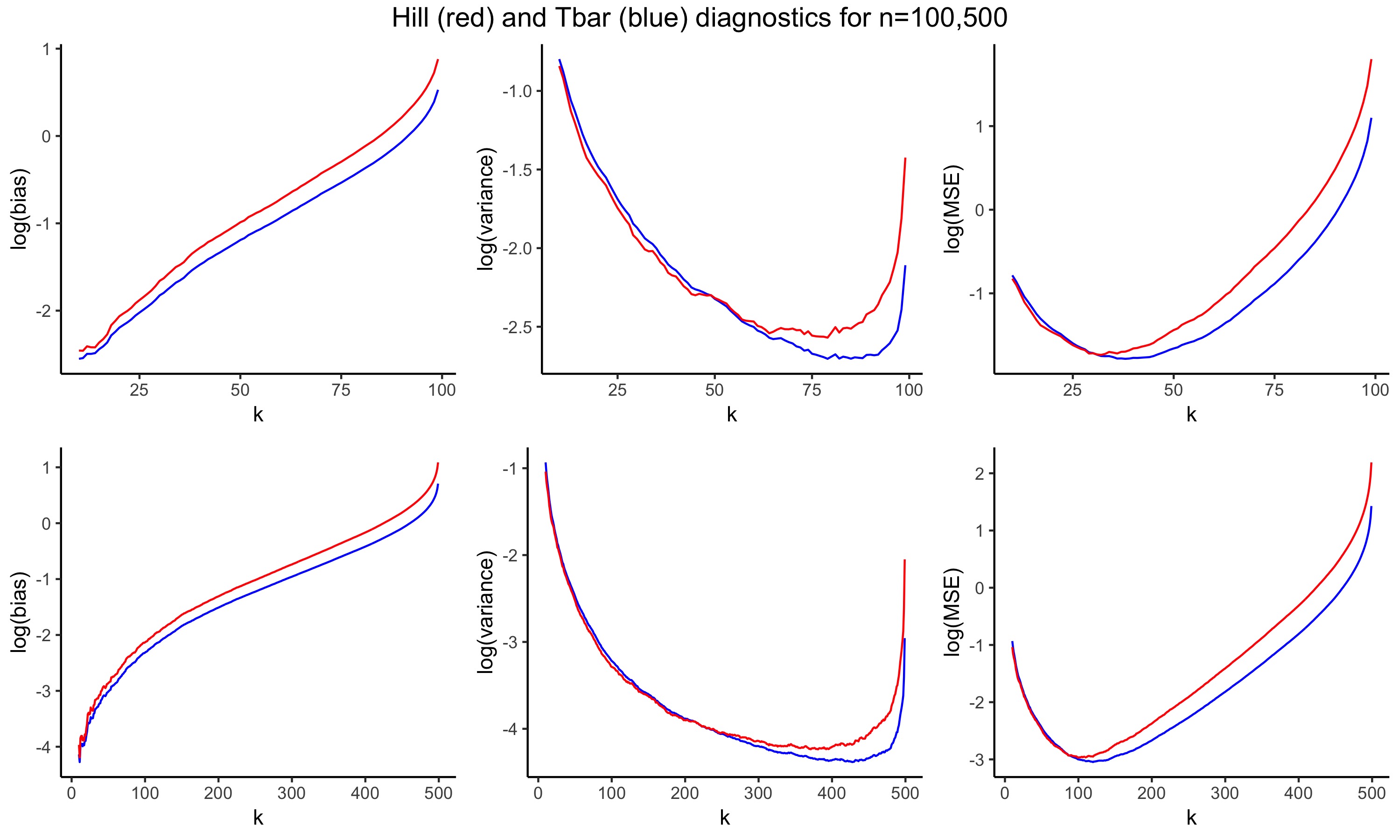}
\caption{Fr\'{e}chet distribution, parameter $\alpha=1/2$. Top: Violin plots for $n=100,500$ of the estimators $H_{\hat k_{GH}}$, $H_{\widehat k^\ast_0,\,p=-1}$, $\overline T_{\hat k_{GH}}$, $\overline T_{\widehat k^\ast_0,\,p=-1}$. Bottom: diagnostics of $\overline T_k$ (blue) and $H_k$ (red) as a function of $k$.} 
\label{frechet_n100_n500_params2}
\end{figure}

\begin{figure}[hh]
\centering
\includegraphics[width=11cm,trim=.5cm 0cm 1.5cm .5cm,clip]{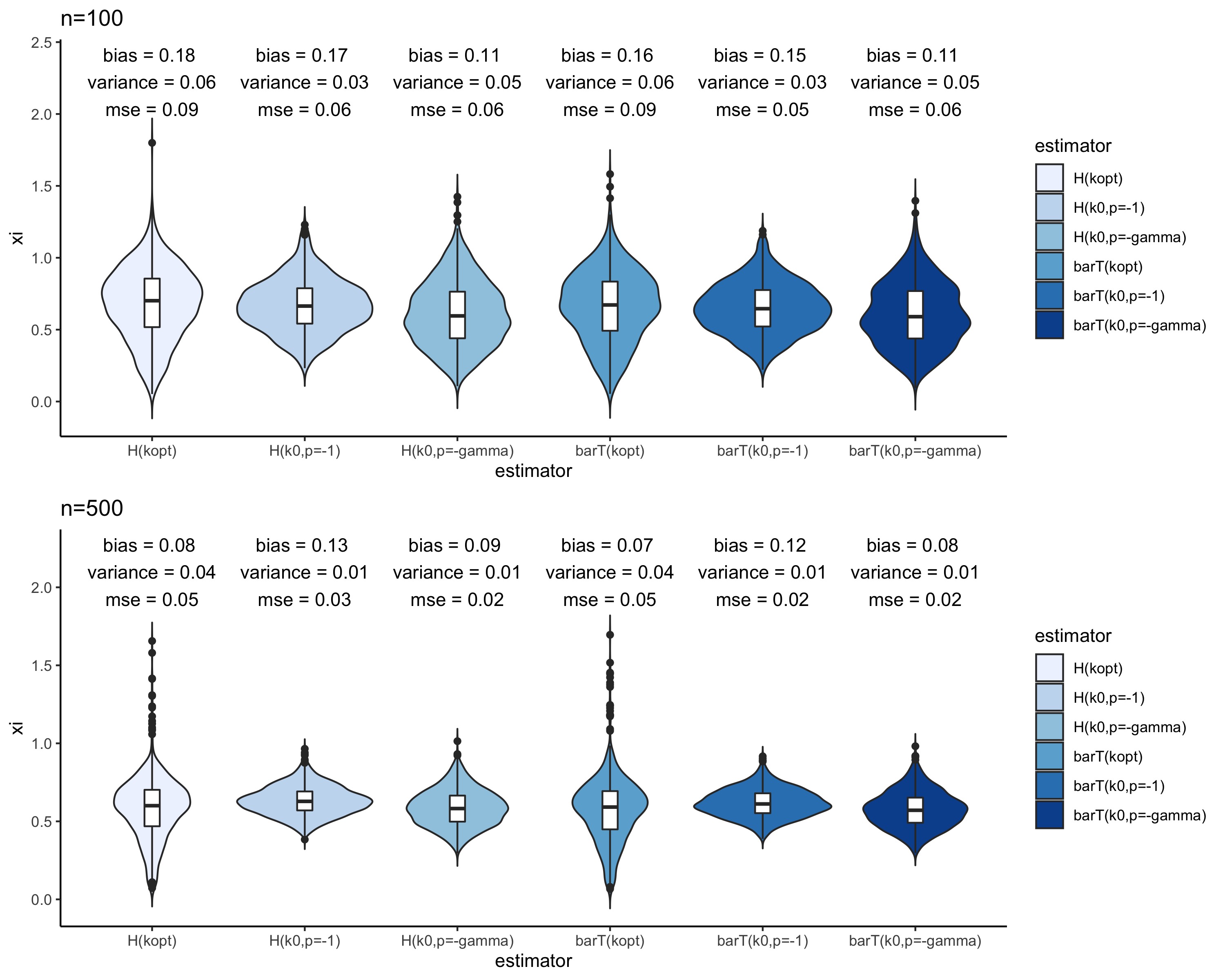}
\includegraphics[width=12cm,trim=.5cm .5cm .5cm 0cm,clip]{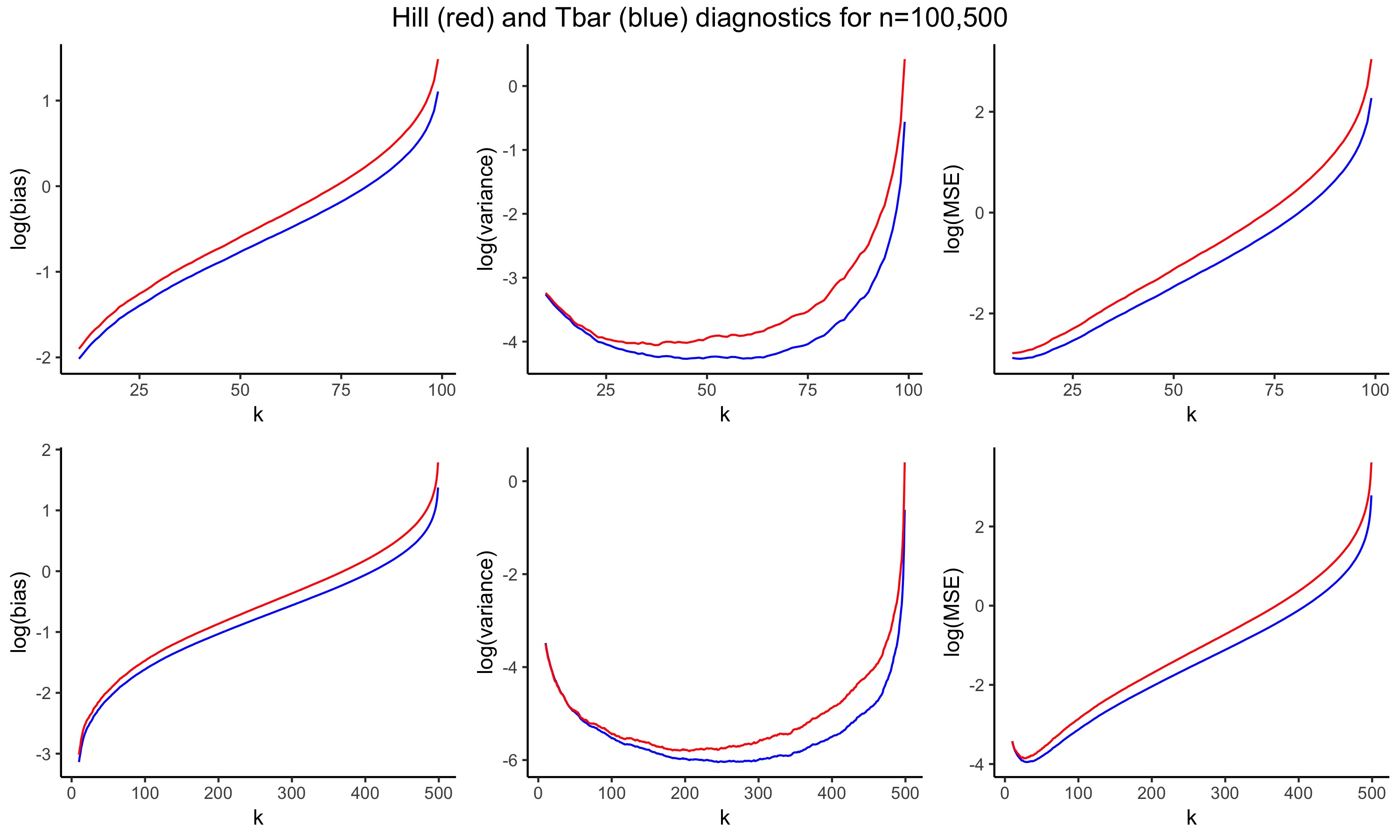}
\caption{GPD distribution, parameters $\gamma=1/2$, $\sigma=2$. Top: Violin plots for $n=100,500$ of the estimators $H_{\hat k_{GH}}$, $H_{\widehat k^\ast_0,\,p=-1}$, $H_{\widehat k^\ast_0,\,p=-\gamma}$, $\overline T_{\hat k_{GH}}$, $\overline T_{\widehat k^\ast_0,\,p=-1}$, $\overline T_{\widehat k^\ast_0,\,p=-\gamma}$. Bottom: diagnostics of $\overline T_k$ (blue) and $H_k$ (red) as a function of $k$.} 
\label{gpd_n100_n500_params1}
\end{figure}

\begin{figure}[hh]
\centering
\includegraphics[width=11cm,trim=.5cm 0cm 1.5cm .5cm,clip]{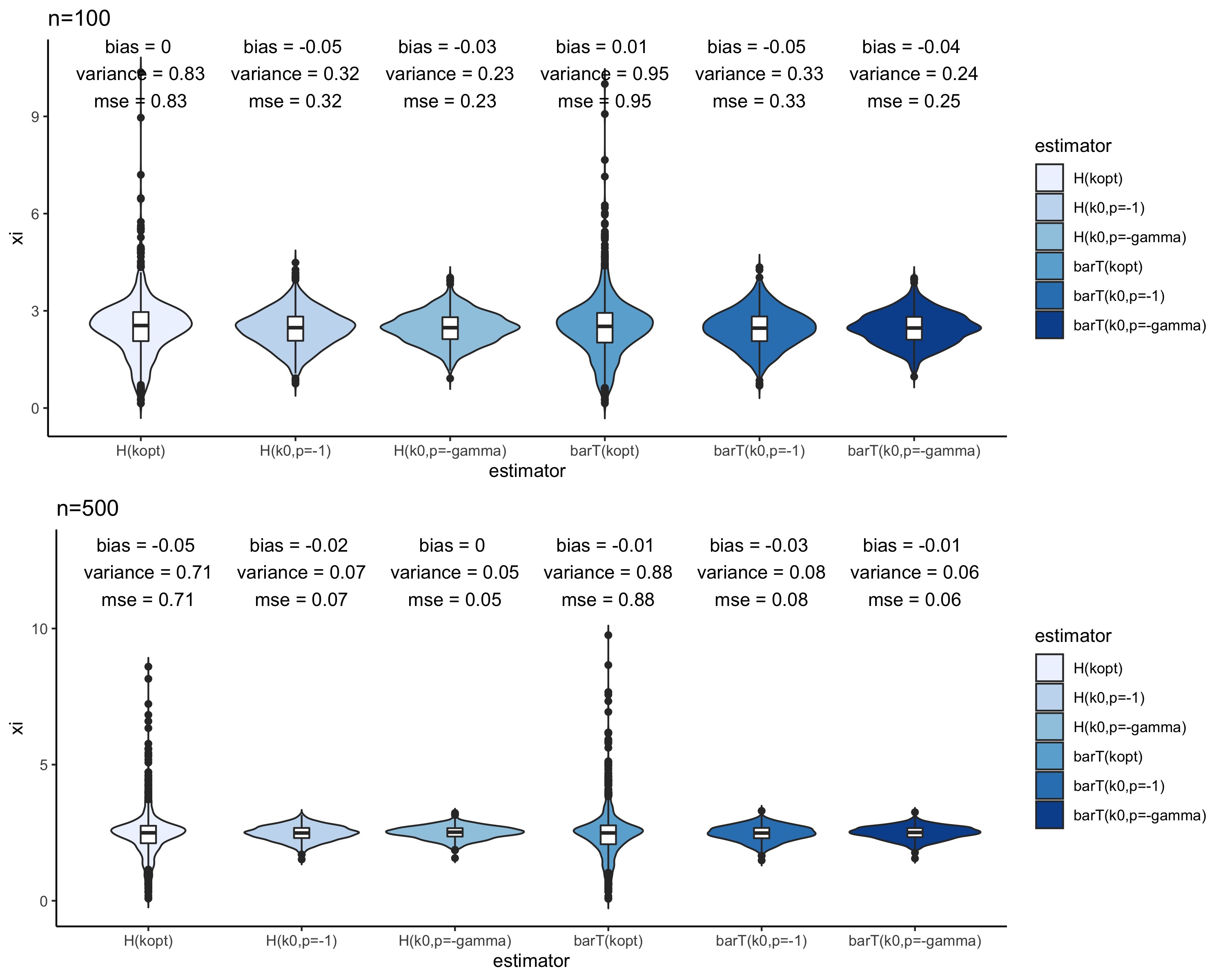}
\includegraphics[width=12cm,trim=.5cm .5cm .5cm 0cm,clip]{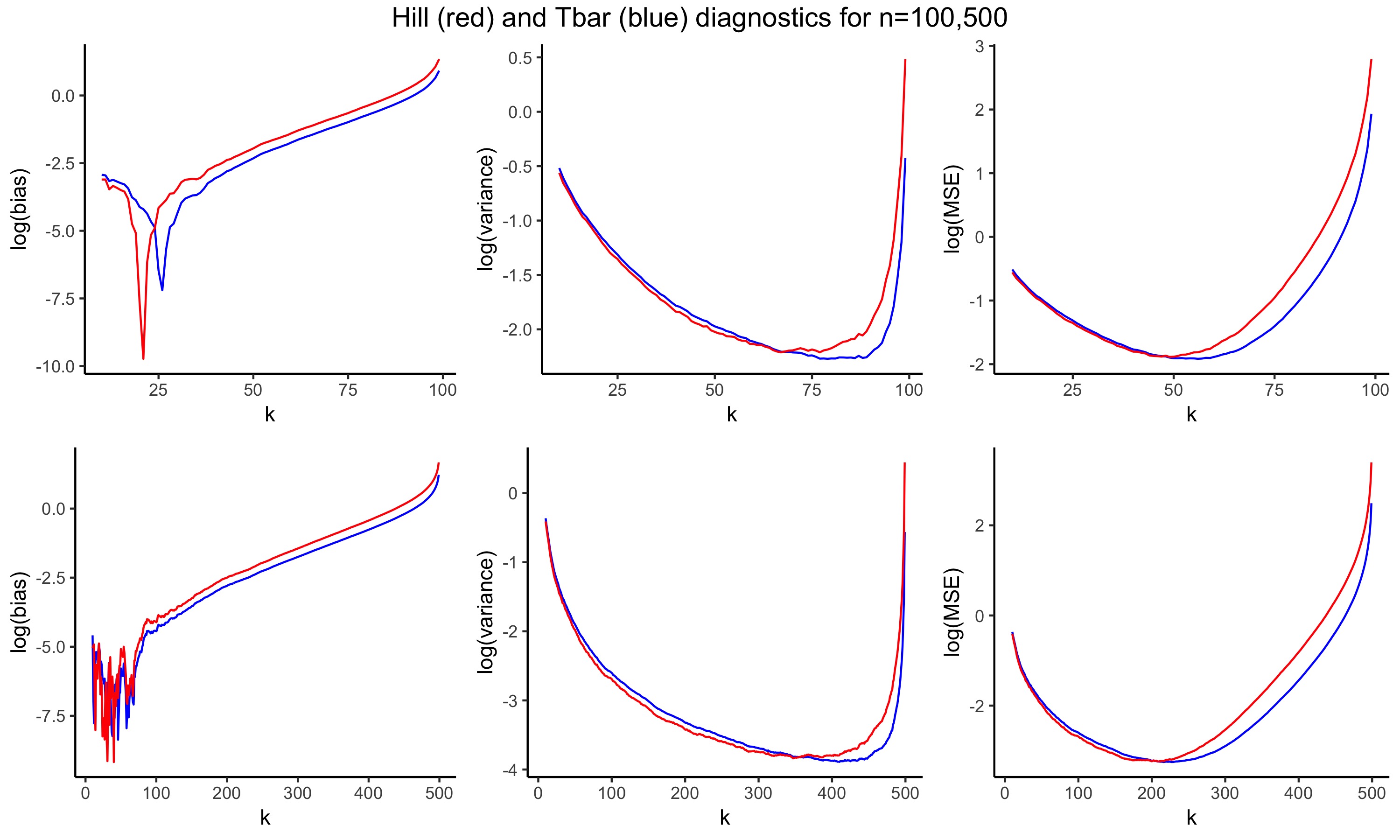}
\caption{GPD distribution, parameters $\gamma=5/2$, $\sigma=1$. Top: Violin plots for $n=100,500$ of the estimators $H_{\hat k_{GH}}$, $H_{\widehat k^\ast_0,\,p=-1}$, $H_{\widehat k^\ast_0,\,p=-\gamma}$, $\overline T_{\hat k_{GH}}$, $\overline T_{\widehat k^\ast_0,\,p=-1}$, $\overline T_{\widehat k^\ast_0,\,p=-\gamma}$. Bottom: diagnostics of $\overline T_k$ (blue) and $H_k$ (red) as a function of $k$.} 
\label{gpd_n100_n500_params2}
\end{figure}

\begin{figure}[hh]
\centering
\includegraphics[width=11cm,trim=.5cm 0cm 1.5cm .5cm,clip]{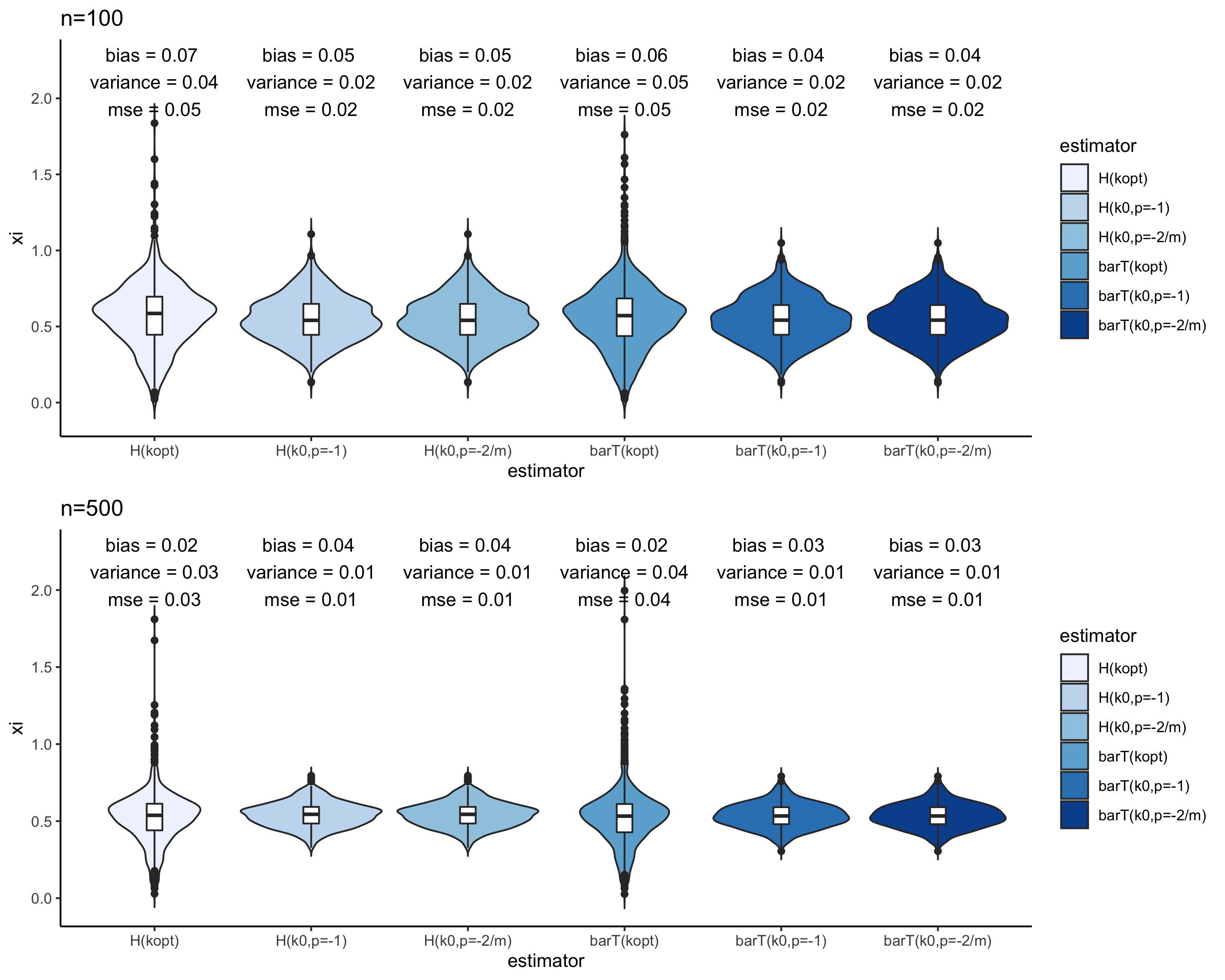}
\includegraphics[width=12cm,trim=.5cm .5cm .5cm 0cm,clip]{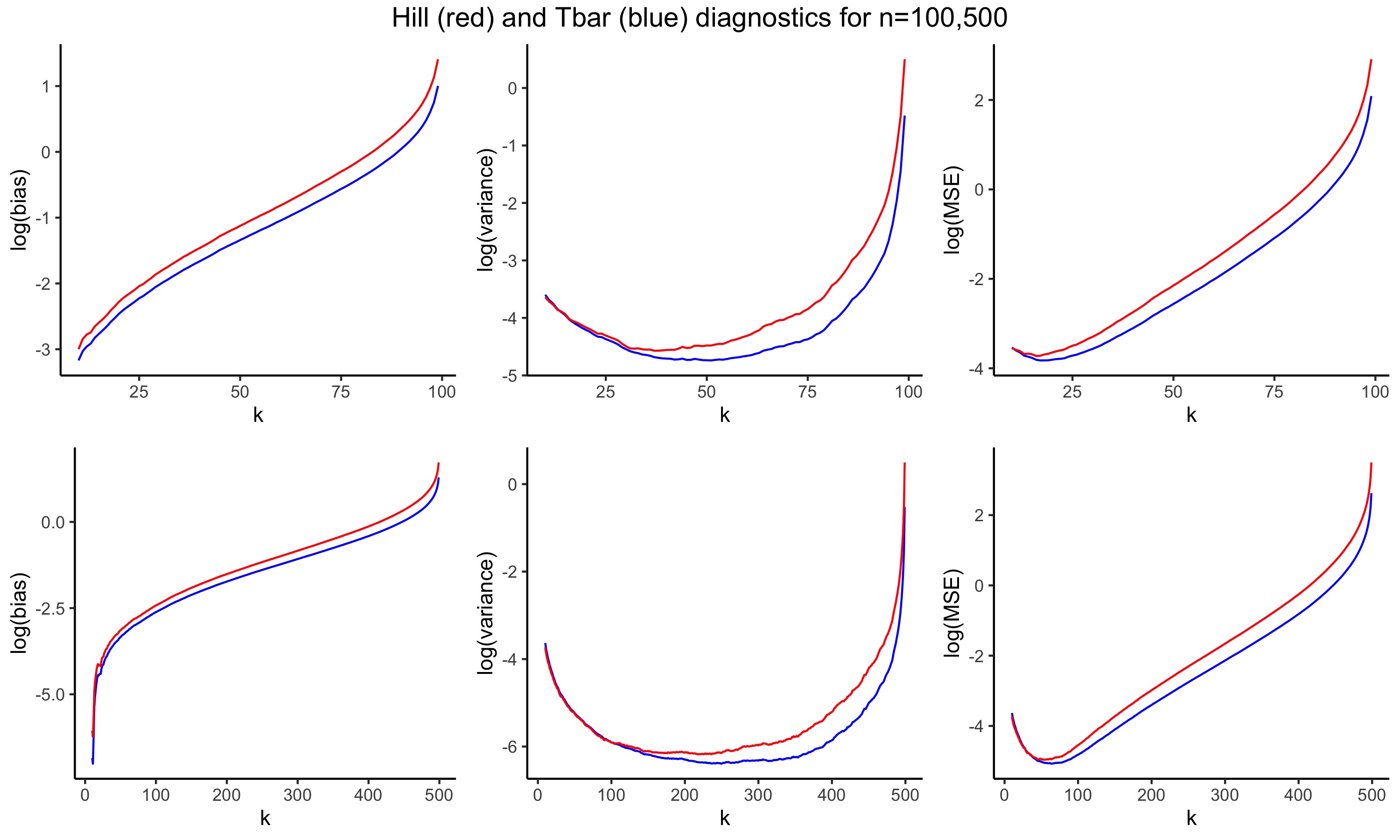}
\caption{Student-t distribution, degrees of freedom $m=2$. Top: Violin plots for $n=100,500$ of the estimators $H_{\hat k_{GH}}$, $H_{\widehat k^\ast_0,\,p=-1}$, $H_{\widehat k^\ast_0,\,p=-2/m}$, $\overline T_{\hat k_{GH}}$, $\overline T_{\widehat k^\ast_0,\,p=-1}$, $\overline T_{\widehat k^\ast_0,\,p=-2/m}$. Bottom: diagnostics of $\overline T_k$ (blue) and $H_k$ (red) as a function of $k$.} 
\label{t_n100_n500_params1}
\end{figure}

\begin{figure}[hh]
\centering
\includegraphics[width=11cm,trim=.5cm 0cm 1.5cm .5cm,clip]{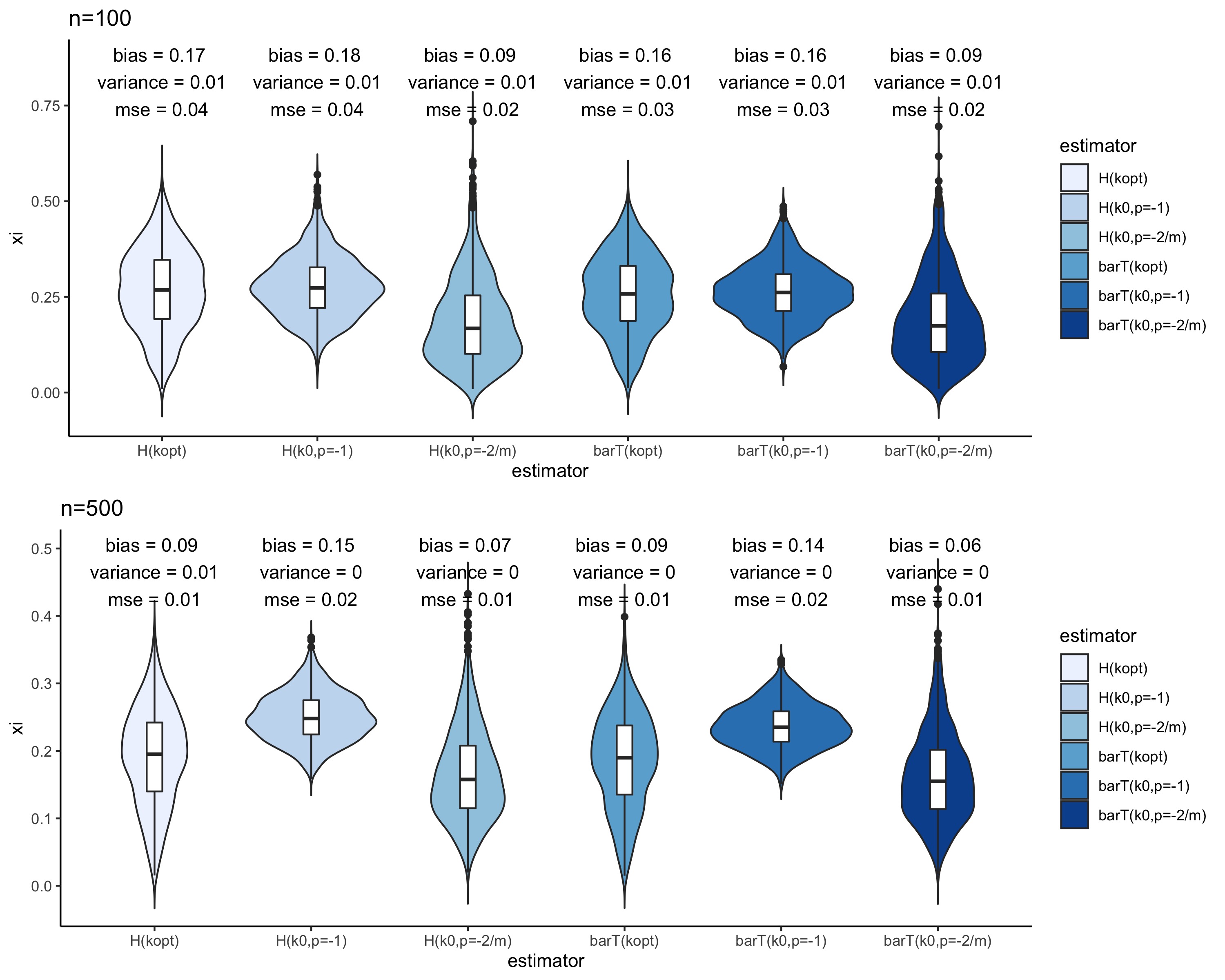}
\includegraphics[width=12cm,trim=.5cm .5cm .5cm 0cm,clip]{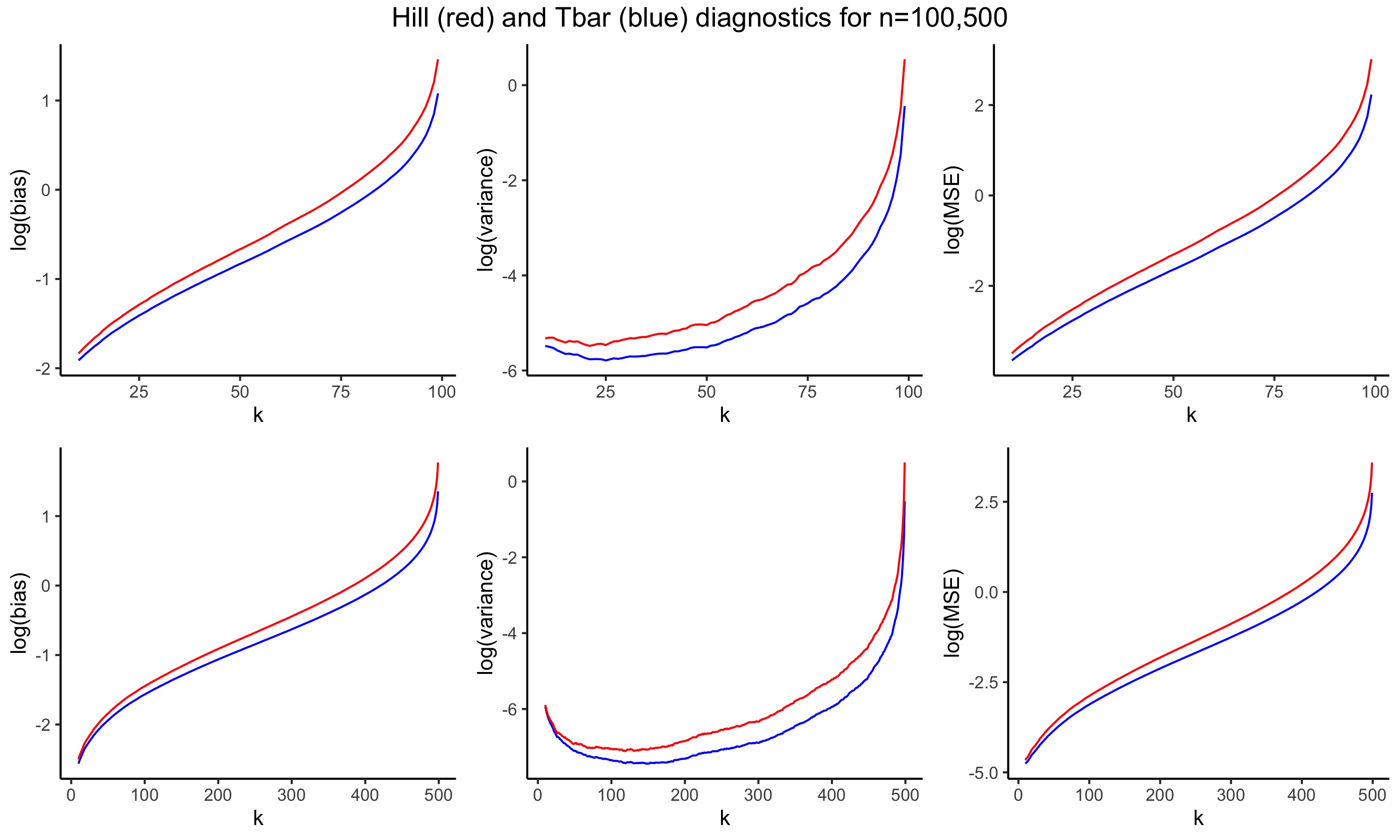}
\caption{Student-t distribution, degrees of freedom $m=10$. Top: Violin plots for $n=100,500$ of the estimators $H_{\hat k_{GH}}$, $H_{\widehat k^\ast_0,\,p=-1}$, $H_{\widehat k^\ast_0,\,p=-2/m}$, $\overline T_{\hat k_{GH}}$, $\overline T_{\widehat k^\ast_0,\,p=-1}$, $\overline T_{\widehat k^\ast_0,\,p=-2/m}$. Bottom: diagnostics of $\overline T_k$ (blue) and $H_k$ (red) as a function of $k$.} 
\label{t_n100_n500_params2}
\end{figure}

\noindent The results are given in Figures \ref{burr_n100_n500_params1}, \ref{burr_n100_n500_params2} for the Burr case; Figures \ref{frechet_n100_n500_params1}, \ref{frechet_n100_n500_params2} for the Fr\'{e}chet case; Figures \ref{gpd_n100_n500_params1}, \ref{gpd_n100_n500_params2} for the GPD case; and Figures \ref{t_n100_n500_params1}, \ref{t_n100_n500_params2} for the Student-t case. We observe that the behaviour is very similar for the four families (which is not uncommon in this context, cf.\ \cite[p.178]{beirlant2002exponential}). 


For the Hill estimator, we notice that our method {  fares well} against the benchmark. The misspecification of the second order parameter $p$ does not play a substantial role, {  except perhaps for the most biased case of the Student-t distribution with $10$ degrees of freedom}. The same behaviour is observed within the three $\overline{T}$-estimators. When comparing Hill against $\overline{T}$-estimators, the latter improve the bias and MSE for nearly all $k$, and in most cases also the variance (except for very heavy tails ($\xi\ge 1$) and small values of $k$). 

Remarkably, the estimator $\overline T_{k_0,\,p=-1}$, where the canonical $p=-1$ is used, is highly competitive against the Hill estimator, especially so for $\xi\le 1$.
This is not a contradiction, since the optimality of the Hill estimator refers to choices for $k$ within the class of $H_{k}$, whereas the $\overline T_k$ estimators span a different class (visible in the weighting interpretation of Section \ref{secw}), and when $k$ is optimized w.r.t.\ AMSE in that class, even better performance can be feasible, which, however, is not the subject of the present paper. 

\section{Insurance data}
\begin{figure}[hh]
\centering
\includegraphics[width=8cm,trim=.5cm 2.5cm 3cm 8cm,clip]{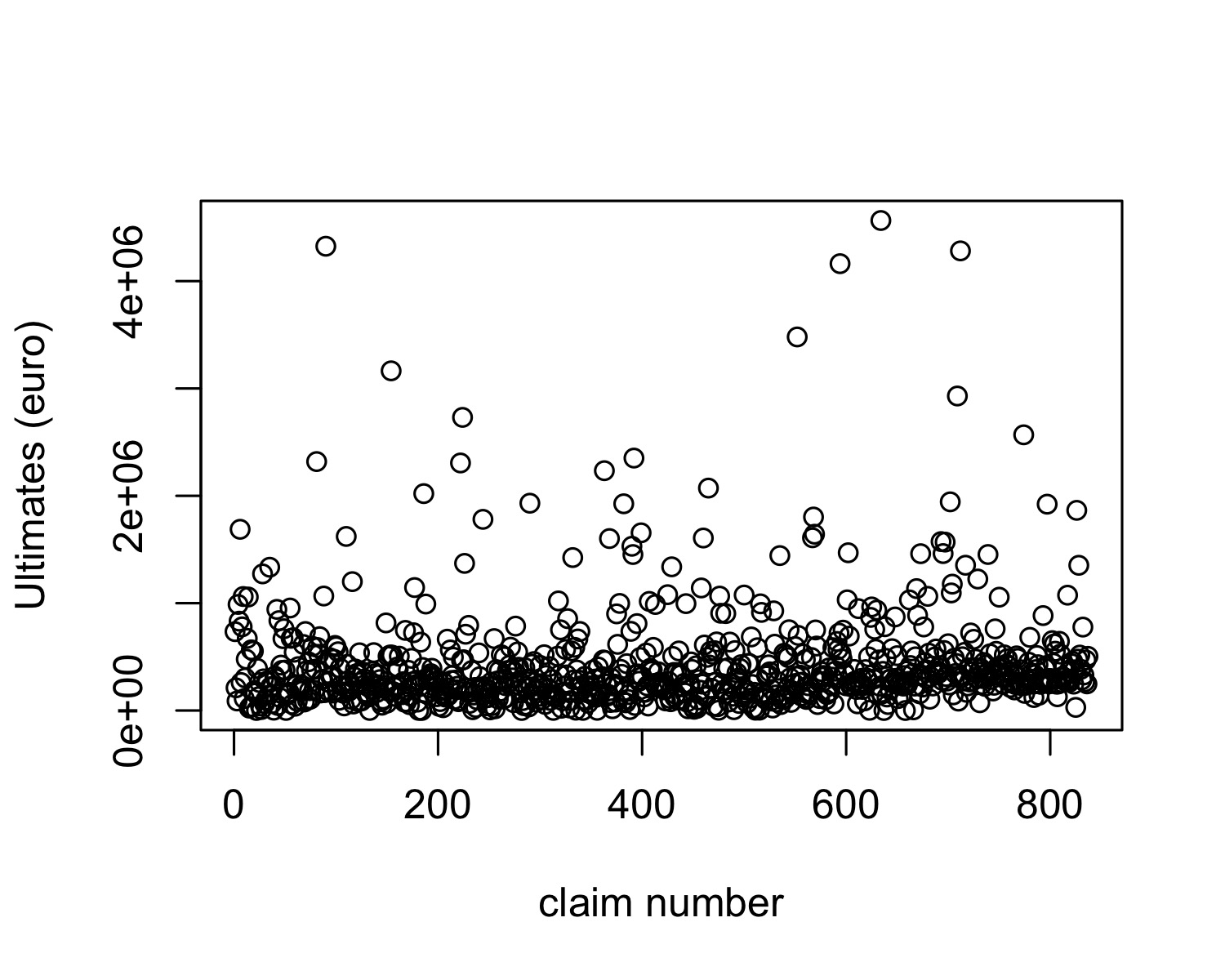}
\caption{Ultimates from an MTPL insurance portfolio.} 
\label{ultimates}
\end{figure}

Let us now consider a real-life insurance data set consisting of $837$ motor third party liability (MTPL) insurance claims from the period 1995-2010 that was studied intensively in \cite{abt} (where it is referred to as "Company A"). These data are right-censored, and were also analyzed recently combining survival analysis techniques and expert information in \cite{combinedtail}. Here, we focus only on the \textit{ultimates}, see Figure \ref{ultimates}, which are the actual final claim sizes for the settled claims and an expert prediction of the total payment until closure for all claims that are still open.

%

\FloatBarrier

\begin{figure}[hh]
\centering
\includegraphics[width=15cm,trim=.5cm .5cm 0cm .5cm,clip]{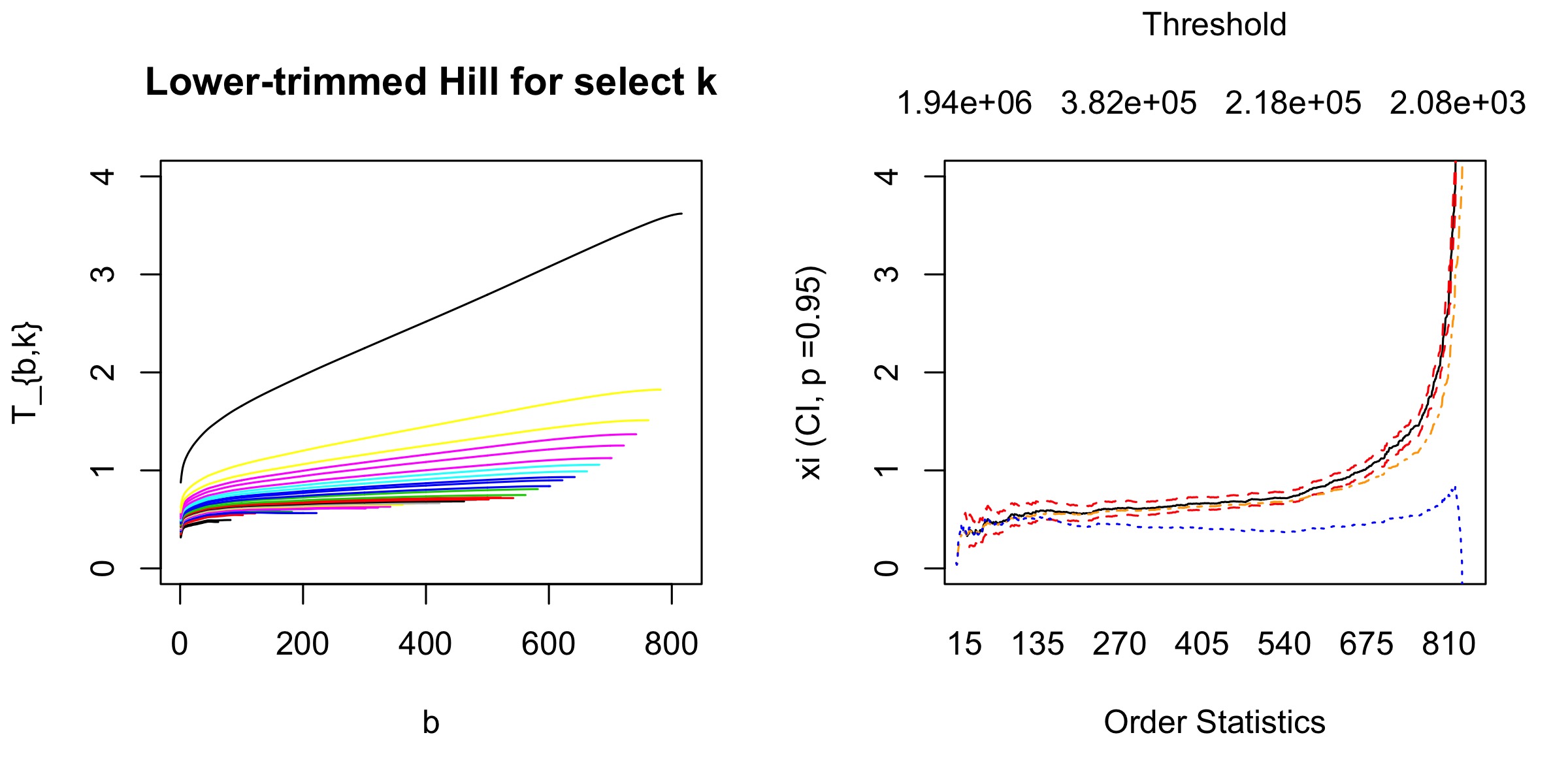}
\includegraphics[width=9cm,trim=.5cm 2cm .5cm 8cm,clip]{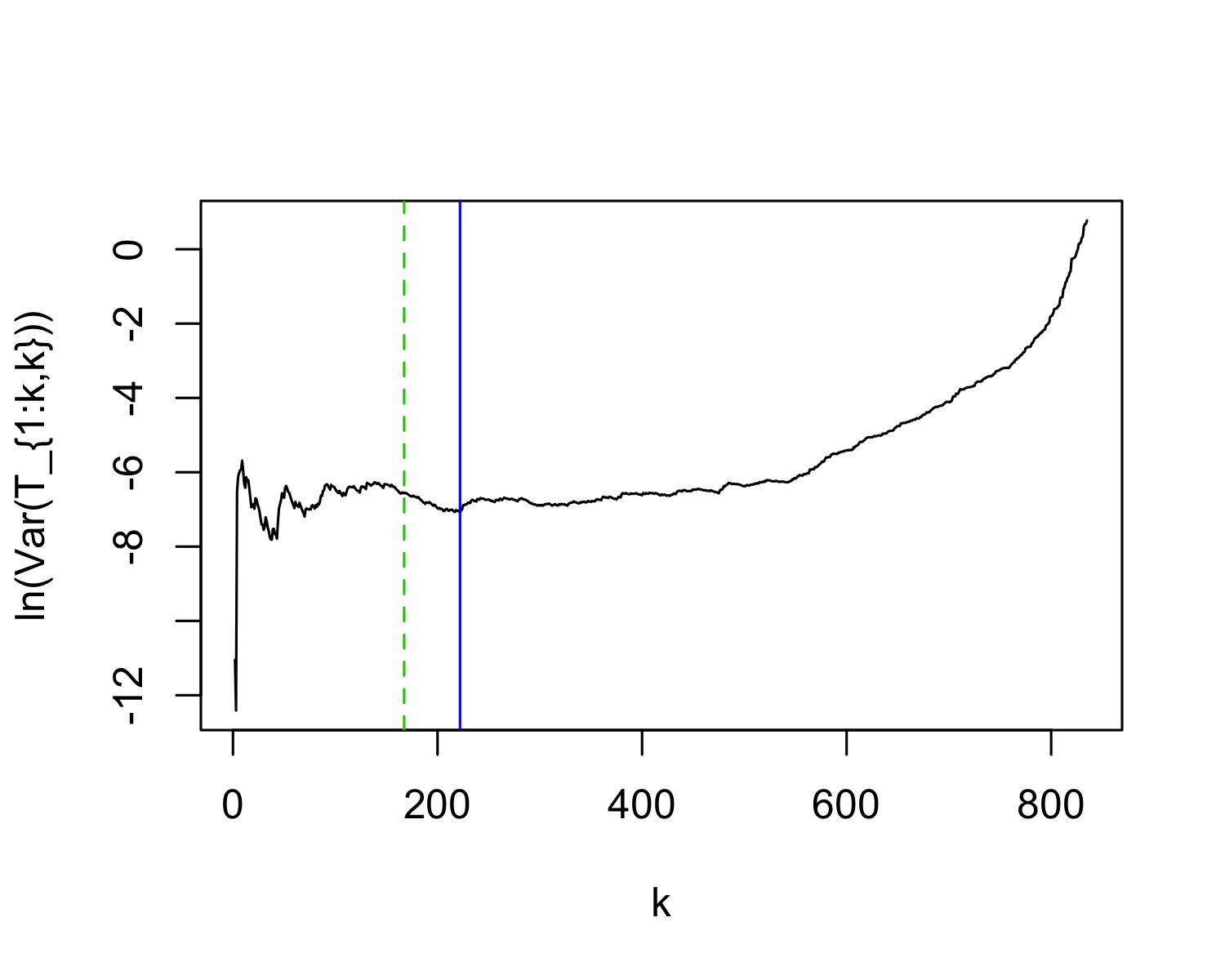}
\caption{MTPL insurance ultimates. Top left: Lower-trimmed Hill estimator (LTH) estimator for varying lower trimming $b$, for $k$ uniformly spaced from 1 to 837 in units of 20. Top right: Hill plot { (black, solid), together with $\overline T_k$ (orange, dashed-and-dotted) and a mean-of-order $-1$ bias-reduced estimator (blue, dotted)}. Bottom: empirical variance of the LTH as a function of $k$. The dotted line is the left limit for  candidates, and the solid line is the resulting minimum.} 
\label{trimmed_ultimates}
\end{figure}

In Figure \ref{trimmed_ultimates} we depict the lower-trimmed Hill plots and the usual Hill plot, together with the empirical variance. {  We have also included $\overline T_k$, which follows somewhat closely the trajectory of the usual Hill plot, and the mean-of-order $-1$ bias-reduced estimator. As a preliminary observation, notice that the $k$-area at which visually the Hill plot and its bias-reduced version start to differ is roughly the same as where the lower trimmed Hill trajectories start flattening out, which serves as a good sanity check.} As in the simulation studies in Section \ref{secsim}, in order to avoid degeneracies, we only look at candidates for the minimizer to the right of $n/5$, which corresponds to $167$ in this case. The minimum empirical variance is then obtained for $\hat{k}^\ast=222$. Using the canonical choice $p=-1$, we have that $\hat{k}^*_0=222/2.62421\approx 85$. Note that for the same choice $p=-1$, using the prior eyeballed estimate $\xi\approx 0.5$, and { based on the Burr-like Hill plot, we can deduce that $\lambda=-1/p=1$, and then $D=1/\tau=-\xi=-0.5$. We thus} get by \eqref{amsemink} { the ad-hoc sample fraction $k^{ah}=112$} (which might be considered the classical choice of the threshold in this case). {  Notice, however, that the latter estimate is only available heuristically, since one needs a first estimate of $\xi$ to estimate $\xi$ itself, and also to make distributional assumptions on the data. We include this estimate here simply as a naive solution that requires no further statistical procedures beyond looking at the Hill plot.}

\begin{figure}[hh]
\centering
\includegraphics[width=16cm,trim=.5cm 2cm .5cm 8cm,clip]{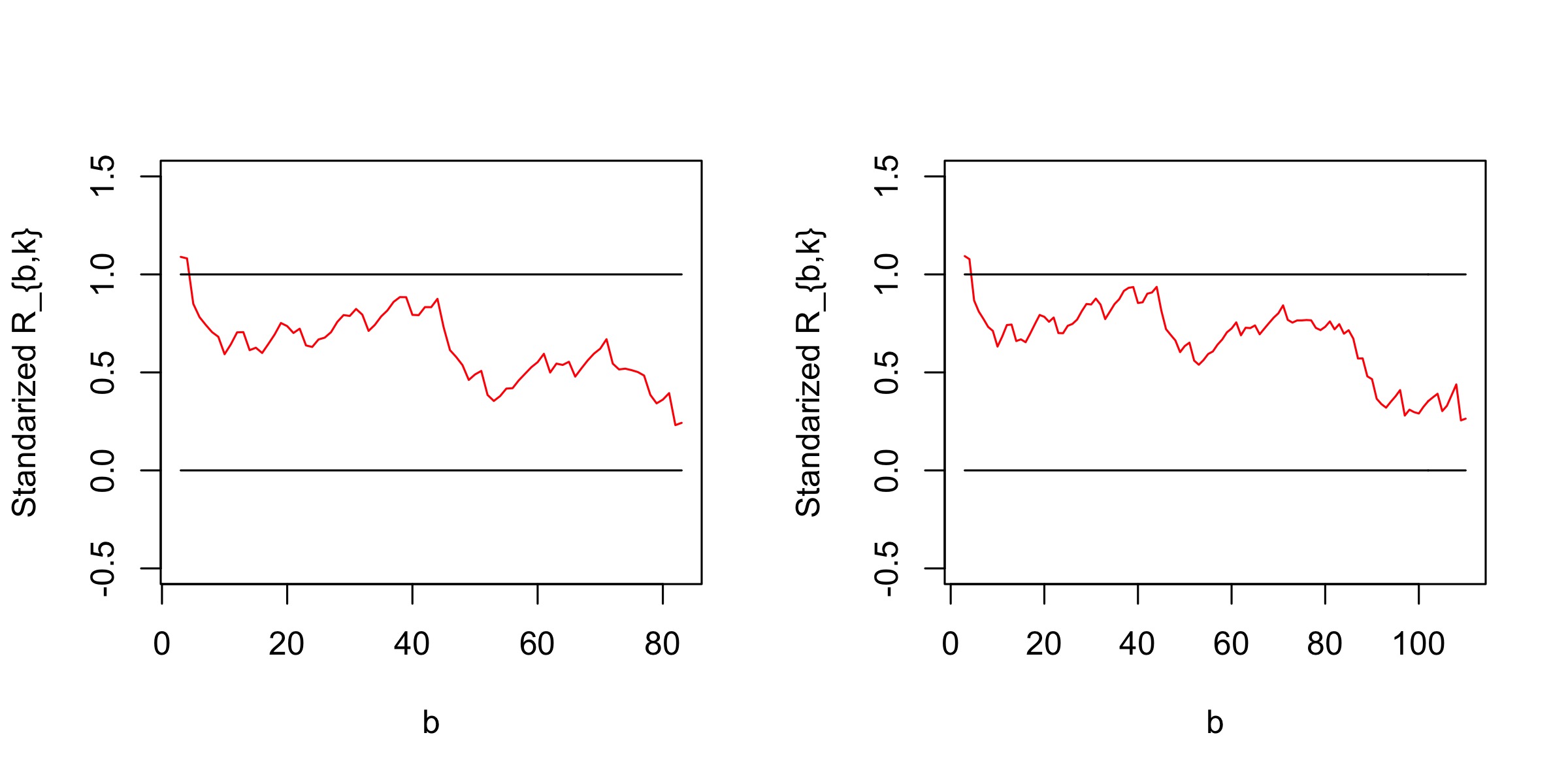}
\caption{Standardized R-statistic for the MTPL ultimates, for the two threshold choices $k=85$ (left) and $k=112$ (right). $N_{MC}=100,000$ and $\alpha=0.05$.} 
\label{rstatultimates}
\end{figure}

\noindent The corresponding estimates of $\xi$ are given by
\begin{align*}
H_{\hat{k}_0^\ast}=0.508,\quad H_{{k}^{ah}}=0.560,\quad \overline T_{\hat{k}_0^\ast}=0.480, \quad \overline T_{{k}^{ah}}=0.525.
\end{align*}
The simulation studies of Section \ref{secsim} may suggest the third of the above numbers to be the most reliable estimate, {  since there is no way of quantifying the eyeball-aided procedure involving $k^{ah}$. However, the $95\%$ confidence interval for the first estimate is $(0.400,\, 0.616)$, suggesting that for a one-sample analysis, it is difficult to make a definitive statement on the statistical superiority of any of the four estimates. Further,} the ratio statistic test in Figure \ref{rstatultimates} suggests that for both thresholds the sample is Pareto in the tail (with only a slight issue for the two largest observations). {  The takeaway is that, roughly speaking, we are able to reconfirm in an automated statistical way what can be deduced by looking at a Hill plot (together with its bias-reduced variants) and guessing the distribution of the sample.}

In \cite[p.99]{abt}, a splicing point was suggested for this data set at around $k=20$, based on expert opinion. A semi-automated option using our method for detecting this splicing point would be to 
replace the left limit $k=167$ by a very small number (in this case $k=4$ is chosen after visual insection of the erratic nature of the empirical variance for the first three), and then to apply our method, which leads to the detection of the minimum variance at $k=38$ (which is clearly visible in Figure \ref{trimmed_ultimates}). Under the assumption $p=-1$ this then leads to $k\approx14$ as a suggested splicing point. {  We would like to point out that the identification of the splicing point matters in insurance practice, since the different resulting  distributional assumptions on either side then have an effect on the location of extreme quantiles and risk management in general, including the capital requirements for solvency purposes. The possibly natural existence of splicing points can also be argued from a causal perspective, as different degrees of inspection scrutiny may be applied below and above certain claim levels.\\
As a side remark, in the present data set the ultimates for the highest claims have a certain degree of intrinsic uncertainty} (as they are just estimates of the final closed claim size), and a more systematic way to approach this particular situation would be to combine the trimming of the Hill estimator from below and above, {  but the latter} is not the focus of the present paper. 
%

\section{Conclusion}
In this paper, we showed that trimming the Hill estimator from the left can lead to favorable properties in connection with the expected empirical variance of the tail index estimators in extreme value statistics. For the Hall class, we established asymptotic results on the behavior of this expected empirical variance, which allows to develop a guideline for the choice of the optimal threshold in the tail index estimation problem. It turns out that there is an intrinsic link between this optimal threshold and the classical optimal threshold for the Hill estimator. Since in the trimming context the identification of the optimal threshold is much more insensitive on the tail characteristics (it only depends on the $p$-parameter in the Hall class, not on $D$ nor on the tail index $\xi$), this link allows to circumvent the classical problem in threshold selection for the Hill estimator. As a by-product, by suitable averaging we develop a novel tail index estimator which assigns a non-uniform weight to each observation in a natural way, relies on fewer assumptions on the tail characteristics, is simple to implement and  outperforms the classical Hill estimator in most cases. The latter is illustrated in simulation studies. In addition, the technique is applied to a real-life insurance data set that was previously studied by other techniques. {  Note also that the proposed selection principle for $k$ based on the variance of the lower-trimmed Hill plots can be applied to any estimator of $\xi$ for which the asymptotic mean squared error can be written as $M_1 {\xi^2 / k} + M_2(p) Q_0 (n/k)^2$ with $M_1>0$ and $M_2(p)>0$  only depending on $p$.} We conclude by noting that the approach taken in this paper is in principle also applicable for the potential improvement of tail index estimators other than the Hill estimator. Further possible directions of future research include the combination of left trimming with right trimming in situations with possible outliers, as well as the consideration of possibly censored data. \\

\FloatBarrier

\textbf{Acknowledgement.} M.B. and H.A. acknowledge financial support from the Swiss National Science Foundation Project 200021\_191984. 

\bibliographystyle{apalike}
\bibliography{Hill_trimming}

\newpage 
\section{Proofs}

\noindent {\bf Proof of Proposition 2.1.}
Set $q=k-b+1$. By the R\'{e}nyi representation \eqref{Renyi_rep},
\begin{align*}
\V\left[T_{b,k}\right]&=\V\left[\sum_{j=1}^k E_j^\ast \sum_{i=j\vee q}^k\frac{\gamma_i}{k-j+1}\right]=\xi^2\;\frac{\sum_{j=1}^k\left(\frac{k-j\vee q+1}{k-j+1}\right)^2}{\left(\sum_{j=1}^k\frac{k-j\vee q+1}{k-j+1}\right)^2}.
\end{align*}
Plugging in $q=1$ ($b=k$) gives 
\begin{align*}
\V\left[T_{k,k}\right]=\frac{\xi^2}{k},
\end{align*}
which corresponds to the usual variance of the Hill estimator $T_{k,k}$ and gives the first identity. In the general case,
\begin{align*}
\V\left[T_{b,k}\right]
&=\xi^2\;\frac{\sum_{j=1}^q\left(\frac{k- q+1}{k-j+1}\right)^2+k-q+1}{\left(\sum_{j=1}^q\frac{k- q+1}{k-j+1}+k-q+1\right)^2}.
\end{align*}
But $j\le q$ implies $\frac{k-q+1}{k-j+1}\le 1$, such that $$\sum_{j=1}^q\frac{1}{k-j+1}\ge \sum_{j=1}^q\frac{k-q+1}{k-j+1}\frac{1}{k-j+1},$$
so 
\begin{align*}
\sum_{j=1}^q\frac{k- q+1}{k-j+1}+k-q+1\ge \sum_{j=1}^q\left(\frac{k- q+1}{k-j+1}\right)^2+k-q+1.
\end{align*}
Thus
\begin{align*}
\V\left[T_{b,k}\right]&\le\xi^2\frac{\sum_{j=1}^q\left(\frac{k- q+1}{k-j+1}\right)^2+k-q+1}{\left(\sum_{j=1}^q\left(\frac{k- q+1}{k-j+1}\right)^2+k-q+1\right)^2}\\
&=\frac{\xi^2}{\sum_{j=1}^q\left(\frac{k- q+1}{k-j+1}\right)^2+k-q+1},
\end{align*}
which gives the second identity.
$\qed$
\\

\noindent
{\bf Proof of Theorem 3.1.}
We first note that
\begin{align*}
T_{b,k}\stackrel{d}{=}\frac{\sum_{i=1}^b\log(U(Y_{n-i+1,n})/U(Y_{n-k,n}))}{b(1+\sum_{j=b+1}^kj^{-1})},
\end{align*}
where $Y_{1,n}<\cdots< Y_{n,n}$ are the order statistics of a standard Pareto sample (the $\xi=1$ case). Then, from the second order condition \eqref{so} we obtain that for $A=Y_{n-k,n}$ and $x=Y_{n-i+1,n}/Y_{n-k,n}$, as $k,n,n/k\to \infty$,
\begin{align*}
T_{b,k}\stackrel{d}{=}\frac{\xi\sum_{i=1}^b \log(Y_{n-i+1,n}/Y_{n-k,n})+\frac{Q_0(Y_{n-k,n})}{p}\sum_{i=1}^b((Y_{n-i+1,n}/Y_{n-k,n})^p-1)(1+o_p(1))}{b(1+\sum_{j=b+1}^kj^{-1})}.
\end{align*}
But by the R\'{e}nyi representation \eqref{Renyi_rep} of exponential order statistics, the first term is distributed as
\begin{align*}
\sum_{i=1}^b\log(Y_{n-i+1,n}/Y_{n-k,n})\stackrel{d}{=}\sum_{j=1}^b E_j+b\sum_{j=b+1}^kE_j/j,
\end{align*}
where $E_1,E_2,\dots, E_k$ are i.i.d.\ standard exponential random variables. For the second term, by convergence to uniform random variables and a Riemann integral approximation, we get
\begin{align*}
\frac 1b \sum_{i=1}^b((Y_{n-i+1,n}/Y_{n-k,n})^p-1)&\stackrel{d}{\approx}\frac 1b \sum_{i=1}^b(((k+1)/i)^p-1)\\
&\approx \frac{k+1}{b}\int_0^{b/(k+1)}(u^{-p}-1)\dd u=\frac{((k+1)/b)^p}{1-p}-1,
\end{align*}
and since $(1-1/Y_{n-k,n})$ is a uniform order statistic, we further get that
\begin{align*}
\frac{Q_0(Y_{n-k,n})}{Q_0(n/k)}\stackrel{P}{\to}1.
\end{align*}
Putting the three pieces together then establishes \eqref{astbk}. $\qed$ \\

\noindent
{\bf Proof of Theorem 3.2.}
With the shortened notation, we write
\begin{align}\label{tbkrep}
T_{b,k}\stackrel{d}{=}\xi \frac{\overline E_b+\sum_{j=b+1}^kE_j/j}{1+\sum_{j=b+1}^kj^{-1}}+Q_0(n/k)c_{b,k,p}(1+o_p(1)),
\end{align}
and by exchange of the order of summation, we can write
\begin{align*}
\overline T_{k}&\stackrel{d}{=}\frac{\xi}{k}  \sum_{b=1}^{k}\frac{\overline E_b+\sum_{j=b+1}^kE_j/j}{1+\sum_{j=b+1}^kj^{-1}}+Q_0(n/k)\overline c_{k,p}(1+o_p(1))\\
&=\frac{\xi}{k}\left[\sum_{j=1}^{k}E_j\sum_{b=j}^{k}\frac{1}{b(1+\sum_{j=b+1}^kj^{-1})}+\sum_{j=2}^k E_j\sum_{b=1}^{j-1}\frac {1}{j(1+\sum_{j=b+1}^kj^{-1})}\right]\\\
&\quad+Q_0(n/k)\overline c_{k,p}(1+o_p(1))\\
&=\frac{\xi}{k}\sum_{j=1}^{k}E_j\left[\sum_{b=j}^{k}\frac{1}{b(1+\log(k/b)}+\sum_{b=1}^{j-1}\frac {1}{j(1+\log(k/b)}\right](1+o(1))\\\
&\quad+Q_0(n/k)\overline c_{k,p}(1+o_p(1)).
\end{align*}
Again, by Riemann integration we have that
\begin{align*}
\frac 1k \sum_{b=j}^{k}\frac{1}{(b/k)(1+\log(k/b))}\approx\int_{j/k}^1\frac{\dd u}{u(1-\log(u))}=\log(1+\log(k/j)),
\end{align*}
and
\begin{align*}
\sum_{b=1}^{j-1}\frac {1}{j(1+\log(k/b)}\approx \frac k j\int_0^{j/k}\frac{\dd u}{1-\log(u)}=\frac {ek}{j} \Ei(1+\log(k/j)).
\end{align*}
 Similarly,
\begin{align}\label{ckpconstant}
\overline c_{k,p}&\approx \frac 1p \int_0^1 \frac{(1-p)^{-1}u^{-p}}{1-\log(u)}\dd u-\frac 1 p \int_0^1\frac {\dd u}{1-\log(u)}\nonumber\\
&=\frac{e^{1-p}}{p(1-p)}\Ei(1-p)-\frac{e}{p}\Ei(1).
\end{align}
Putting the pieces together then indeed yields \eqref{tkrep}.
$\qed$
\\

\noindent
{\bf Proof of Theorem 3.3.}
Let us first decompose each summand by writing
\begin{align*}
\E[(T_{b,k}-\overline T_k)^2]=\E[(T_{b,k}-\xi)^2]+\E[(\overline T_{k}-\xi)^2]-2\E[(T_{b,k}-\xi)(\overline T_{k}-\xi)],
\end{align*}
and subsequently consider each term separately. From \eqref{tbkrep} we have that
\begin{align*}
\E[(T_{b,k}-\xi)^2]&=\V[T_{b,k}]+\mbox{Bias}^2[T_{b,k}]
&=\xi^2 \frac{\frac{1}{b}+\sum_{j=b+1}^k1/j^2}{(1+\sum_{j=b+1}^kj^{-1})^2}+Q_0^2(n/k)c_{b,k,p}^2(1+o_p(1)).
\end{align*}
On the other hand, \eqref{tkrep} gives
\begin{align*}
\E[(\overline T_{k}-\xi)^2]&=\V[\overline T_{k}]+\mbox{Bias}^2[\overline T_{k}]\\
&=\frac{\xi^2}{k^2}\sum_{j=1}^{k}\left[\log(1+\log(k/j))+\frac {ek}{j} \Ei(1+\log(k/j))\right]^2(1+o(1))\\
&\quad+Q_0^2(n/k)\left[\frac{e^{1-p}}{p(1-p)}\Ei(1-p)-\frac{e}{p}\Ei(1)\right]^2(1+o_p(1)).
\end{align*}
The third term can be analyzed using both \eqref{tkrep} and \eqref{tbkrep} as follows, 
\begin{align*}
&\E[(T_{b,k}-\xi)(\overline T_{k}-\xi)]=\E[(T_{b,k}-\E[T_{b,k}])(\overline T_{k}-\E[\overline T_{k}])]+Q_0^2(n/k)c_{b,k,p}\overline c_{k,p}\\
&=\xi^2\E\left[\left(\frac{\frac 1 b \sum_{j=1}^b (E_j-1)+\sum_{j=b+1}^k(E_j-1)/j}{1+\sum_{j=b+1}^kj^{-1}}\right) \left(\sum_{i=1}^k\frac {E_i -1}{k} S(i,k)(1+o(1))\right)\right]\\
&\quad+Q_0^2(n/k)c_{b,k,p}\overline c_{k,p}\\
&=\xi^2\frac{\sum_{j=1}^k (j\vee b)^{-1} S(j,k)}{k(1+\sum_{j=b+1}^kj^{-1})}(1+o(1))+Q_0^2(n/k)c_{b,k,p}\overline c_{k,p}
\end{align*}
where $S(j,k):=\log(1+\log(k/j))+\frac {ek}{j} \Ei(1+\log(k/j))$.\\

We now proceed to add the $k$ summands of the expected variance. To this end, some preparatory calculations will be helpful. By \eqref{cbkpconstant} and Riemann approximation we have
\begin{align*}
\frac{1}{k}\sum_{b=1}^{k}c_{b,k,p}^2&\approx \frac{1}{k}\sum_{b=1}^{k}\frac{1}{p^2}\cdot \frac{\frac{((k+1)/b)^{2p}}{(1-p)^2}-2\frac{((k+1)/b)^p}{1-p}+1}{(1+\log((k+1)/b))^2}\\
&\approx \frac{1}{p^2(1-p)^2}\left[1-e^{1-2p}(1-2p)\Ei(1-2p)\right]\\
&\quad -\frac{2}{p^2(1-p)}\left[1-e^{1-p}(1-p)\Ei(1-p)\right]\\
&\quad +\frac{1}{p^2}\left[1-e\Ei(1)\right].
\end{align*}
By virtue of \eqref{ckpconstant},
\begin{align*}
\overline c_{k,p}^2 \approx \frac{e^{2(1-p)}}{p^2(1-p)^2}\Ei^2(1-p)-2 \frac{e^{2-p}}{p^2(1-p)}\Ei(1-p)\Ei(1)+\frac{e^2}{p^2}\Ei^2(1),
\end{align*}
from which we deduce that as $k\to \infty$,
\begin{align*}
\frac{1}{k}\sum_{b=1}^{k}c_{b,k,p}^2-\overline c_{k,p}^2\to f(p),
\end{align*}
where $f(p)$ is given by \eqref{fp}.

Observe that {  for the terms arising from the expected covariance term}
\begin{align*}
\frac{1}{k}\sum_{b=1}^{k}\frac{\frac{1}{b}+\sum_{j=b+1}^k1/j^2}{(1+\sum_{j=b+1}^kj^{-1})^2}&\approx\frac{2}{k}\int_0^1\frac{\dd u}{u(1-\log(u))^2}-\frac{1}{k}\int_0^1\frac{\dd u}{(1-\log(u))^2}\\
&=\frac{1+e\Ei(1)}{k}.
\end{align*}

\noindent Next,
\begin{align*}
&\frac{1}{e}\frac{1}{k}\sum_{b=1}^{k}\frac{\sum_{j=1}^b b^{-1} (\log(1+\log(k/j))+\frac {ek}{j} { \Ei}(1+\log(k/j)))}{1+\log(k/b)}\\
&\approx \int_0^1\frac{1}{z\log(e/z)}\left(\int_0^z \frac{1}{u} \left(\int_{\log(e/u)}^\infty\log(v)e^{-v}\dd v\right)\dd u\right) \dd z\\
&\approx 0.266=:I_1
\end{align*}
and
\begin{align*}
&\frac{1}{e}\frac{1}{k}\sum_{b=1}^{k}\frac{\sum_{j=b+1}^k j^{-1} (\log(1+\log(k/j))+\frac {ek}{j} {  \Ei}(1+\log(k/j)))}{1+\log(k/b)}\\
&\approx \int_0^1\frac{1}{\log(e/z)}\left(\int_z^1 \frac{1}{u^2} \left(\int_{\log(e/u)}^\infty\log(v)e^{-v}\dd v\right)\dd u\right) \dd z\\
&\approx0.135746=:I_2.
\end{align*}
Finally, { for the mean squared error term of $\overline{T}_k$ we find}
\begin{align*} 
&\frac{1}{k}\sum_{j=1}^{k}(k/j)^2\left(\int_{\log(e/u)}^\infty \log(v)e^{-v}\dd v\right)^2\\
&\approx\int_0^1u^{-2}\left(\int_{\log(e/u)}^\infty \log(v)e^{-v}\dd v\right)^2 \dd u\\
&\approx 0.148005=:I_3.
\end{align*}
Altogether we hence obtain
\begin{align*}
&\E\left[\frac{1}{k}\sum_{b=1}^{k}(T_{b,k}-\overline T_k)^2\right]\\
&=\frac{1}{k}\sum_{b=1}^{k}\left(\E[(T_{b,k}-\xi)^2]+\E[(\overline T_{k}-\xi)^2]-2\E[(T_{b,k}-\xi)(\overline T_{k}-\xi)]\right)\\
&=\xi^2\left(\frac{1+e\Ei (1)}{k}\right)(1+o(1))+\xi^2\frac{e^2}{k}I_3(1+o(1))\\
&-2\xi^2 \frac{e}{k}(I_1+I_2)(1+o(1))+Q_0^2(n/k)\left[\frac{1}{k}\sum_{b=1}^{k}c_{b,k,p}^2-\overline c_{k,p}^2\right](1+o_p(1))\\
&=\xi^2\left(\frac{1+e\Ei(1)}{k}\right)(1+o(1))+\xi^2\frac{e^2}{k}I_3(1+o(1))\\
&-2\xi^2 \frac{e}{k}(I_1+I_2)(1+o(1))+Q_0^2(n/k)f(p)(1+o_p(1))\\
&=\frac{C}{k} \xi^2(1+o(1))+Q_0^2(n/k)f(p)(1+o_p(1))\,
\end{align*}
with
\begin{align*}
C=1+e\Ei(1)+e^2I_3-2e(I_1+I_2)\approx 0.502727.
\end{align*}
\hfill $\qed$

\end{document}